\documentclass[sigconf, nonacm, 10pt, screen]{acmart}

\AtBeginDocument{%
  }
\usepackage{amsmath,amsfonts}
\usepackage{algorithmic}
\usepackage{algorithm}
\usepackage{array}
\usepackage{textcomp}
\usepackage{stfloats}
\usepackage{url}
\usepackage{verbatim}
\usepackage{graphicx}
\usepackage{hyperref}
\usepackage{booktabs}
\usepackage{multirow}
\usepackage{threeparttable}
\usepackage{xcolor}
\usepackage{colortbl}
\usepackage{makecell}
\usepackage{tabularx}
\usepackage{wasysym} 
\usepackage{xcolor}       
\usepackage{pifont}       
\usepackage{ragged2e}
\usepackage{enumitem}
\usepackage{appendix}
\usepackage{soul}
\usepackage{multicol}

\setlist[itemize]{noitemsep, topsep=0pt}
\setlist[enumerate]{noitemsep, topsep=0pt}
\setlist[itemize]{leftmargin=0.4cm} 
\setlist[enumerate]{leftmargin=0.4cm} 
\definecolor{lightblue}{RGB}{135,206,250}
\definecolor{lightred}{RGB}{255,182,193}
\newcommand{\cmark}{\textcolor{green!60!black}{\ding{51}}}  
\newcommand{\xmark}{\textcolor{red}{\ding{55}}}            
\setcopyright{none} 
\usepackage{caption}
\makeatletter
\renewcommand\section{\@startsection{section}{1}{0pt}%
  {1.5ex plus 0.2ex minus 0.1ex}
  {0.8ex plus 0.1ex}
  {\normalfont\Large\bfseries}}

\renewcommand\subsection{\@startsection{subsection}{2}{0pt}%
  {1.2ex plus 0.2ex minus 0.1ex}
  {0.6ex plus 0.1ex}
  {\normalfont\large\bfseries}}

\renewcommand\subsubsection{\@startsection{subsubsection}{3}{0pt}%
  {1ex plus 0.1ex minus 0.1ex}
  {0.5ex plus 0.1ex}
  {\normalfont\normalsize\bfseries}}
\makeatother
\pagebreak[1]

\begin{document}
\title{WiLLM: an Open Framework for LLM Services over Wireless Systems}
\author{Boyi Liu}
\affiliation{%
  \institution{HKUST \& UCL}
  \country{}
}
\email{bliubd@connect.ust.hk}

\author{Yongguang Lu}
\affiliation{%
  \institution{SYSU \& PCL}
  \country{} 
}
\email{luyg5@mail2.sysu.edu.cn}

\author{Jianguo Zhao}
\affiliation{%
  \institution{BUPT \& PCL}
  \country{}
}
\email{zhaojianguo@bupt.edu.cn}

\author{Qiang Yang}
\affiliation{%
  \institution{University of Cambridge}
  \country{}
}
\email{qy258@cam.ac.uk}

\author{Wen Wu}
\affiliation{%
  \institution{Pengcheng Laboratory}
  \country{}
}
\email{wuw02@pcl.ac.cn}

\author{Lin Chen}
\affiliation{%
  \institution{Macao Polytechnic University}
  \country{}
}
\email{lchen@mpu.edu.mo}

\author{Jagmohan Chauhan}
\affiliation{%
  \institution{University College London}
  \country{}
}
\email{jagmohan.chauhan@ucl.ac.uk
}

\author{Jun Zhang}
\affiliation{%
  \institution{HKUST}
  \country{}
}
\email{eejzhang@ust.hk}
\begin{abstract}
Large Language Model (LLM) services fundamentally differ from traditional Deep Neural Network (DNN) applications in wireless networks. We identify three critical distinctions: (1) unlike traditional DNNs with unidirectional data flows, LLM's multimodal interactions create bidirectional heavy loads with contrasting bottlenecks, requiring direction-aware resource scheduling; (2) while traditional DNNs exhibit fixed computational patterns, LLM's highly variable inference times interact complexly with network slicing, causing dynamic bottleneck migration; and (3) in contrast to predictable DNN traffic, LLM's token streams demonstrate unprecedented burstiness and state dependencies. These insights motivate WiLLM, the first open-source framework, implemented as a wireless platform, for LLM service research. Built on OpenAirInterface, WiLLM introduces several technical innovations: dynamic slice compatibility, universal UE compatibility through application-layer tunneling, multi-UE multi-slice scheduling, dual-mode resource allocation, and cross-layer APIs. In addition, WiLLM eliminates the need for specialized wireless expertise, enabling researchers and developers to experiment with LLM services over realistic cellular networks. We demonstrate the platform's capabilities through a smart glasses case study and provide a comprehensive dataset of \~1.6 million synchronized measurements. 
The complete system, dataset, and appendix are available at \url{https://openwillm.github.io}.
\end{abstract}

\maketitle
\addtolength{\floatsep}{-\baselineskip}
\addtolength{\dblfloatsep}{-\baselineskip}
\addtolength{\textfloatsep}{-\baselineskip}
\addtolength{\dbltextfloatsep}{-\baselineskip}
\addtolength{\abovedisplayskip}{-1ex}
\addtolength{\belowdisplayskip}{-1ex}
\addtolength{\abovedisplayshortskip}{-1ex}
\addtolength{\belowdisplayshortskip}{-1ex}
\section{Introduction}
\label{introduction}
The proliferation of Large Language Models across mobile applications creates unprecedented demands on wireless network infrastructure. From augmented reality assistants to real-time translation services, LLM-powered applications require seamless integration of computational inference and wireless communication~\cite{bariah2024large}. However, deploying LLM services over cellular networks reveals fundamental mismatches between their operational characteristics and existing network architectures~\cite{boateng2024survey, chaoub2025mobile}. Modern wireless systems, evolved to handle video streaming and web traffic, struggle with the unique patterns of token generation, multimodal processing, and bidirectional heavy loads that characterize LLM interactions.
\begin{figure}[!htpb]
    \centering
    \vspace{15pt}
    \includegraphics[width=0.48\textwidth]{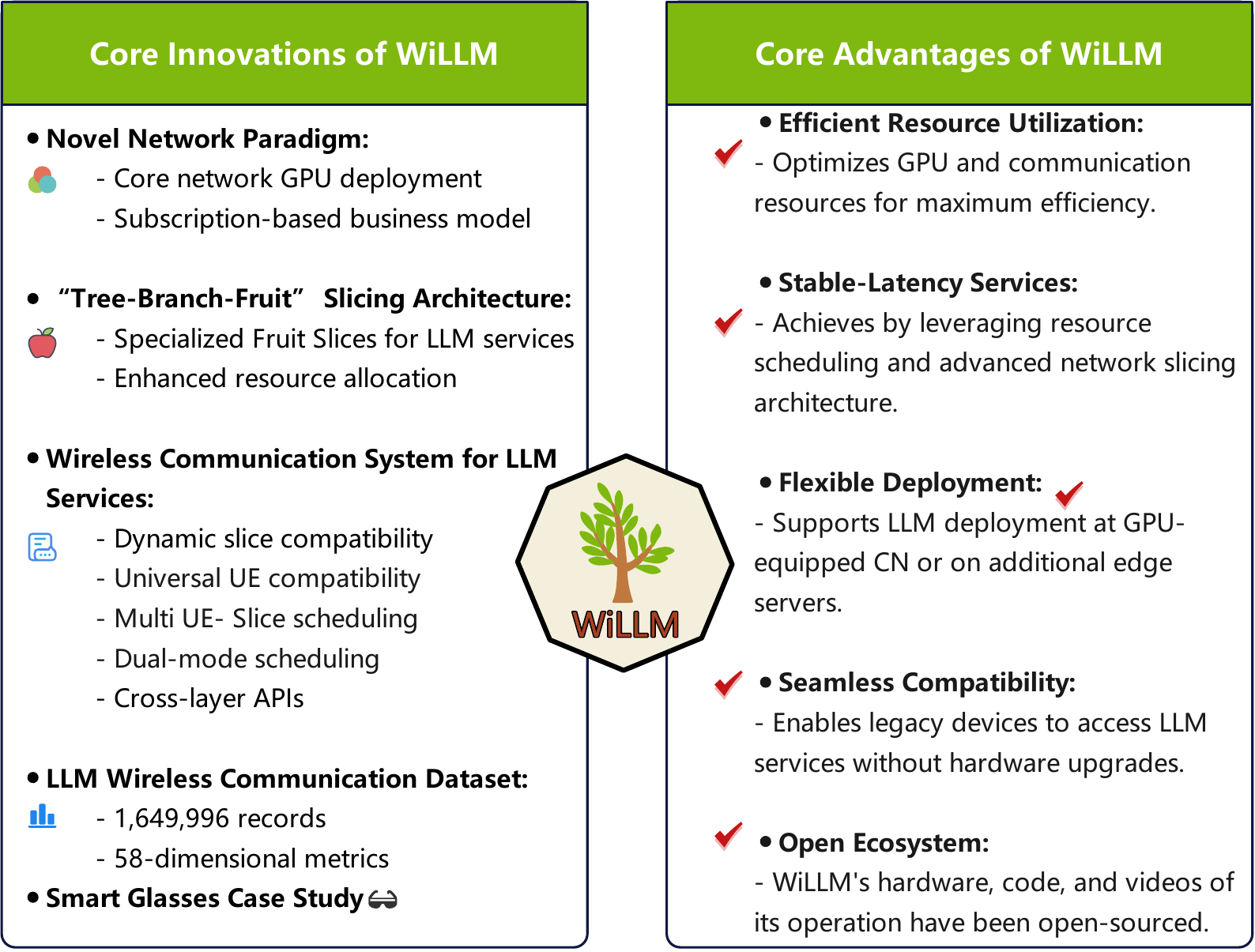}
    \caption{Contributions and advantages of the proposed WiLLM.}
    \vspace{10pt}
    \label{fig:firstFigure}
\end{figure}

\begin{figure}[!htbp]
    \centering
    \includegraphics[width=0.48\textwidth]{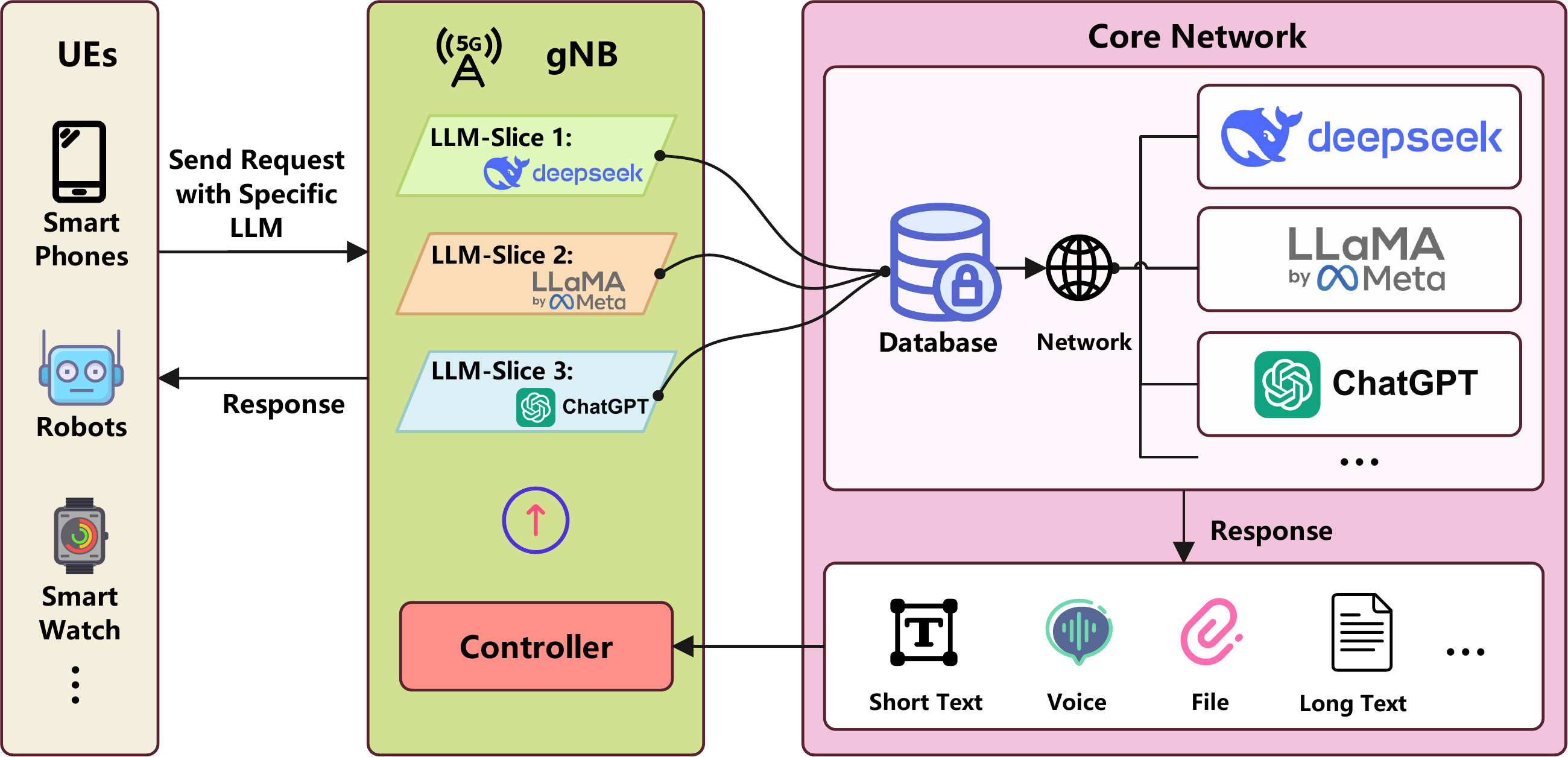}
    \caption{The LLM Services Pipeline over WiLLM~\cite{liu2024llm}.}
    \label{fig:slice_function_architecture}
\end{figure}
The challenge extends beyond simple capacity constraints. Our measurements across diverse deployment scenarios reveal that LLM services violate core assumptions underlying current wireless system designs. Traditional traffic engineering approaches assume predictable patterns, unidirectional dominance, and static computational requirements~\cite{andrade2025study}. LLM services break these assumptions through their generative nature, creating workloads that dynamically shift between computation and communication bottlenecks within single sessions. This fundamental mismatch prevents effective service delivery and optimization using existing tools and methodologies.

Compounding these technical challenges, the research ecosystem lacks accessible platforms for studying LLM behavior in wireless environments. Current open source testbeds such as OpenAirInterface~\cite{nikaein2014openairinterface} and srsRAN~\cite{gomez2016srslte} require deep telecommunications expertise and specialized hardware, effectively excluding general researchers and developers from wireless experimentation. Meanwhile, LLM deployment frameworks like Ollama~\cite{ollama} and TensorRT-LLM~\cite{tensorrt_llm} operate without awareness of wireless constraints. This divide between communities impedes innovation at their intersection, where the most impactful advances will emerge in mobile AI services~\cite{zheng2024review}.

Our extensive empirical analysis, detailed in Section~\ref{sec:characteristics}, identifies specific characteristics that distinguish LLM services from traditional applications and motivate specialized system design. These findings drive WiLLM, an open-source platform that bridges the gap between wireless infrastructure and LLM service. Rather than treating LLM services as generic data traffic, WiLLM incorporates architectural innovations specifically addressing their unique requirements while maintaining accessibility for researchers without wireless expertise.

Figure~\ref{fig:firstFigure} and~\ref{fig:slice_function_architecture} illustrates WiLLM's advantages and the service pipeline. Built on OpenAirInterface's robust protocol implementation, WiLLM extends the base platform with LLM-aware capabilities while preserving compatibility with standard devices and networks. The system enables researchers to experiment with real cellular networks using commercial hardware, deploy various LLM models from Llama~\cite{llama_cpp} to LLaVA~\cite{liu2023llava}, and observe detailed cross-layer interactions through comprehensive monitoring interfaces. These capabilities transform wireless LLM research from a specialized domain requiring extensive telecommunications knowledge to an accessible field where AI researchers can directly contribute innovations. To evaluate the platform's capability, we conducted a case study with smart glasses, where a team with minimal wireless expertise successfully deployed and optimized a multimodal LLM service. This validation confirms that WiLLM achieves its dual objectives: providing the technical capabilities necessary for LLM service delivery while removing barriers that have historically separated wireless and AI research communities.

This paper makes three primary contributions. (1) We provide empirical evidence establishing how LLM services differ fundamentally from traditional network applications, based on analysis of over 1.6 million synchronized measurements across network layers. (2) We present WiLLM as an open experimental platform that democratizes wireless LLM research through architectural innovations and intuitive interfaces. (3) We release our complete implementation, comprehensive dataset, and benchmarking tools to foster an ecosystem for wireless LLM innovation.

The remainder of this paper is organized as follows. Section~\ref{sec:characteristics} analyzes the unique characteristics of LLM services through empirical measurements. Section~\ref{sec:architecture} presents WiLLM's system architecture and design principles. Section~\ref{sec:implementation} details our implementation on OpenAirInterface. Section~\ref{sec:evaluation} validates our design through comprehensive evaluation. Section~\ref{sec:related} reviews related work, and Section~\ref{sec:conclusion} concludes.
\vspace{-2pt}
\section{LLM Services: Why Different from Traditional DNN over Wireless Systems?}
\label{sec:characteristics}
We deployed WiLLM to systematically characterize LLM service behavior across commercial UE devices, a 5G RAN with configurable resource scheduling, and a GPU-accelerated Core Network. Our experimental design combines static and dynamic configurations for both UE and gNB components across four operational scenarios, isolating the effects of system adaptability and network slicing on service performance. Multimodal workloads include image-to-text queries that stress uplink capacity, and text-to-image generation that creates diverse downlink requirements. NTP-based synchronization achieves millisecond-level accuracy for cross-layer correlation, enabling us to trace requests from application initiation through network transmission to inference completion. Through careful control of network conditions and computational isolation, we collected extensive empirical evidence revealing how LLM services fundamentally differ from traditional wireless applications, as detailed in the following three insights.
\subsection{Insight 1: Bidirectional Heavy Load Asymmetry}
\label{subsec:insight1}
Traditional DNN applications in wireless networks exhibit clear unidirectional patterns. For example, image classification services upload raw images and receive lightweight labels, while video analytics applications stream uplink data for cloud processing. This unidirectional dominance enables straightforward optimization: allocate resources primarily for the dominant direction.

\begin{figure*}[t]
    \centering
    \includegraphics[width=1\textwidth]{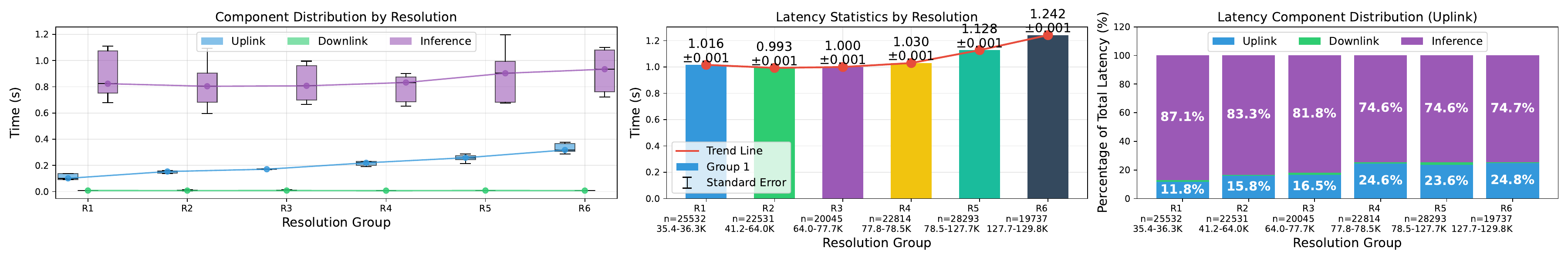}
    \caption{Component distribution and latency analysis for uplink transmissions across resolution groups. The x-axis labels R1-R6 represent various resolution groups for image requests sent to LLM, where R1-R6 correspond to increasing resolutions. The left panel shows the latency composition of different components (uplink, downlink, inference), the center panel presents statistical trends with standard errors, and the right panel quantifies each component's percentage contribution to total latency.}
    \label{fig:Uplink_UE_Dynamic}
\end{figure*}

\begin{figure*}[t]
    \centering
    \includegraphics[width=1\textwidth]{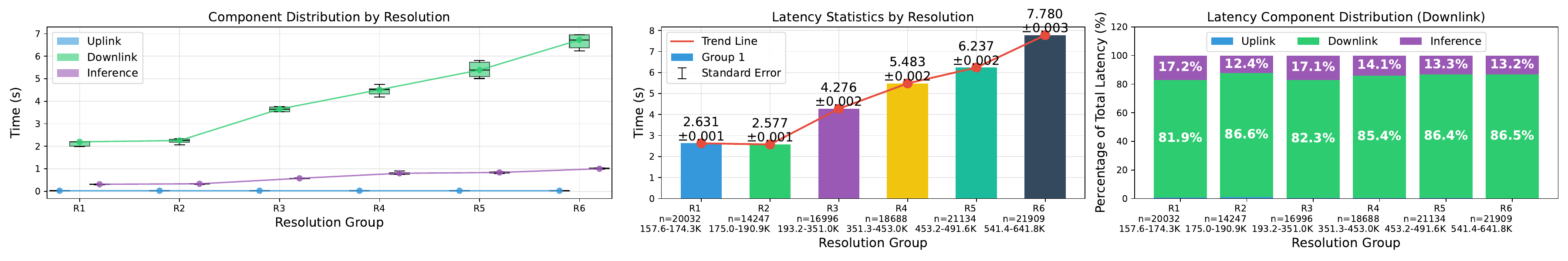} 
    \caption{Component distribution and latency analysis for downlink transmissions across resolution groups. Similar to Figure~\ref{fig:Uplink_UE_Dynamic}, it provides three complementary views: component distribution, statistical latency metrics, and proportional component analysis for downlink scenarios.}
    \label{fig:Downlink_UE_Dynamic}
\end{figure*}

LLM services fundamentally break this pattern due to their multimodal interaction capabilities. Within a single session, users seamlessly transition between uploading images for scene understanding and requesting image generation from text descriptions. Our measurements reveal that this creates two contrasting operational modes with opposite bottlenecks. As shown in Figure~\ref{fig:Uplink_UE_Dynamic}, in uplink-heavy scenarios (image-to-text), computational inference at the LLM dominates the end-to-end latency, accounting for 74\% to 87\% of total delay, while network transmission contributes marginally. Conversely, Figure~\ref{fig:Downlink_UE_Dynamic} demonstrates that in downlink-heavy scenarios (text-to-image), network transmission becomes the primary constraint, consuming 81\% to 86\% of total latency, with inference time reduced from 12\% to 17\%.

This asymmetry stems from the fundamental nature of multimodal LLMs. Processing high-resolution images requires extensive computation for feature extraction and semantic understanding, creating uplink inference bottlenecks. Meanwhile, generating detailed visual content produces large data payloads that stress downlink capacity. The continuous nature of LLM interactions means both patterns occur within the same session, making traditional unidirectional optimization strategies ineffective. Systems must dynamically recognize and adapt to these contrasting requirements, a capability absent in current wireless architectures designed for predictable traffic patterns.

\subsection{Insight 2: Computation-Communication Coupling in Slicing}
\label{subsec:insight2}
Traditional DNN workloads demonstrate predictable computational patterns. A ResNet-50~\cite{kaiming2015resnet} model requires approximately 4 GFLOPs per image regardless of network conditions. A YOLO~\cite{redmon2016you} detector maintains consistent inference time across deployments. This computational predictability decouples inference from network resource allocation, enabling independent optimization of computation and communication resources.
\begin{figure}[t]
    \centering
    \includegraphics[width=0.48\textwidth]{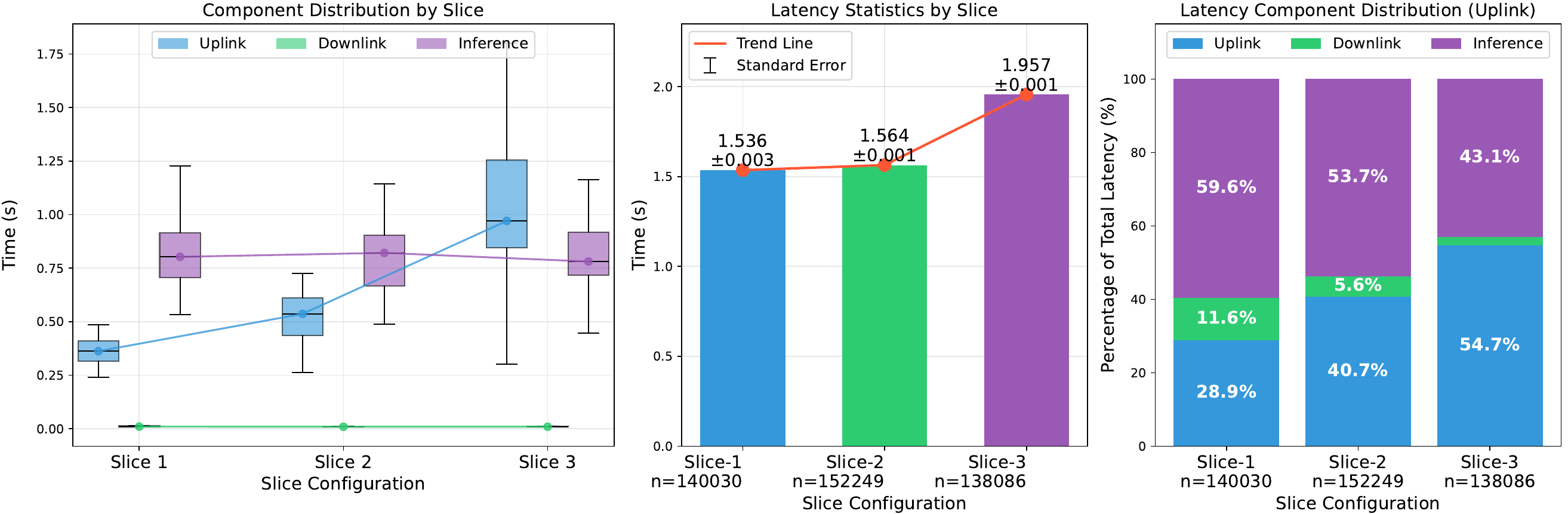}
    \caption{Component distribution and latency analysis across slice groups. Communication resources for uplink slices in Slice 1 to Slice 3 are gradually increasing. The middle panel statistically validates performance differences, while the rightmost panel quantifies how slice configurations shift processing time proportions among uplink, downlink, and inference operations.}
    \label{fig:Uplink_gNB_Dynamic}
\end{figure}

LLM services exhibit fundamentally different behavior where computational demands vary dramatically and interact complexly with network resources. Unlike traditional DNNs with fixed architectures processing fixed-size inputs, LLMs generate variable-length outputs through autoregressive decoding. The inference time depends on multiple dynamic factors, such as the input and output token count, sampling temperature, beam search width, accumulated context length, and more. Our measurements reveal that identical prompts can result in inference times varying by over 300\% based on these generation parameters.

More critically, this computational variability interacts with network slicing to create dynamic bottleneck migration. As shown in Figure~\ref{fig:Uplink_gNB_Dynamic}, when network slice resources increase from minimal to abundant allocation, the system bottleneck shifts dramatically. Under constrained slice configurations, inference time dominates at 59.6\% of total latency, as limited network resources throttle the overall pipeline. However, with enhanced slice resource allocation, the bottleneck migrates to transmission, which grows to 54.7\% of latency. This migration occurs because improved network resources enable faster data transfer, exposing computational limitations that were previously masked by network constraints.

This coupling between computation and communication invalidates traditional slice optimization approaches that assume fixed service characteristics. LLM services require dynamic adaptation mechanisms that consider both computational state and network conditions, adjusting resource allocation based on real-time bottleneck identification rather than static service profiles.
\subsection{Insight 3: Token Stream Unpredictability}
\label{subsec:insight3}
Traditional DNN applications generate highly predictable network traffic. Classification models produce fixed-size outputs~\cite{krizhevsky2017imagenet}; object detectors generate bounded detection lists~\cite{zhao2019object, redmon2016you}; even video streams maintain consistent bitrates through rate control~\cite{om2022h, alsmirat2023video}. This predictability enables efficient resource allocation through statistical multiplexing and quality-of-service guarantees based on well-characterized traffic models.

LLM token streams violate these assumptions through three unique characteristics~\cite{stojkovic2025dynamollm}. First, the generative nature creates inherent unpredictability in output length. A simple query may yield trigger a terse response or an elaborate explanation, depending on subtle prompt variations or sampling randomness. Second, token generation exhibits extreme burstiness as the model alternates between rapid token transmission and computational pauses for attention calculation. Third, multi-turn interactions introduce state dependencies where previous exchanges influence subsequent resource requirements through context accumulation.

\begin{figure*}[t]
    \centering
    \includegraphics[width=1\textwidth]{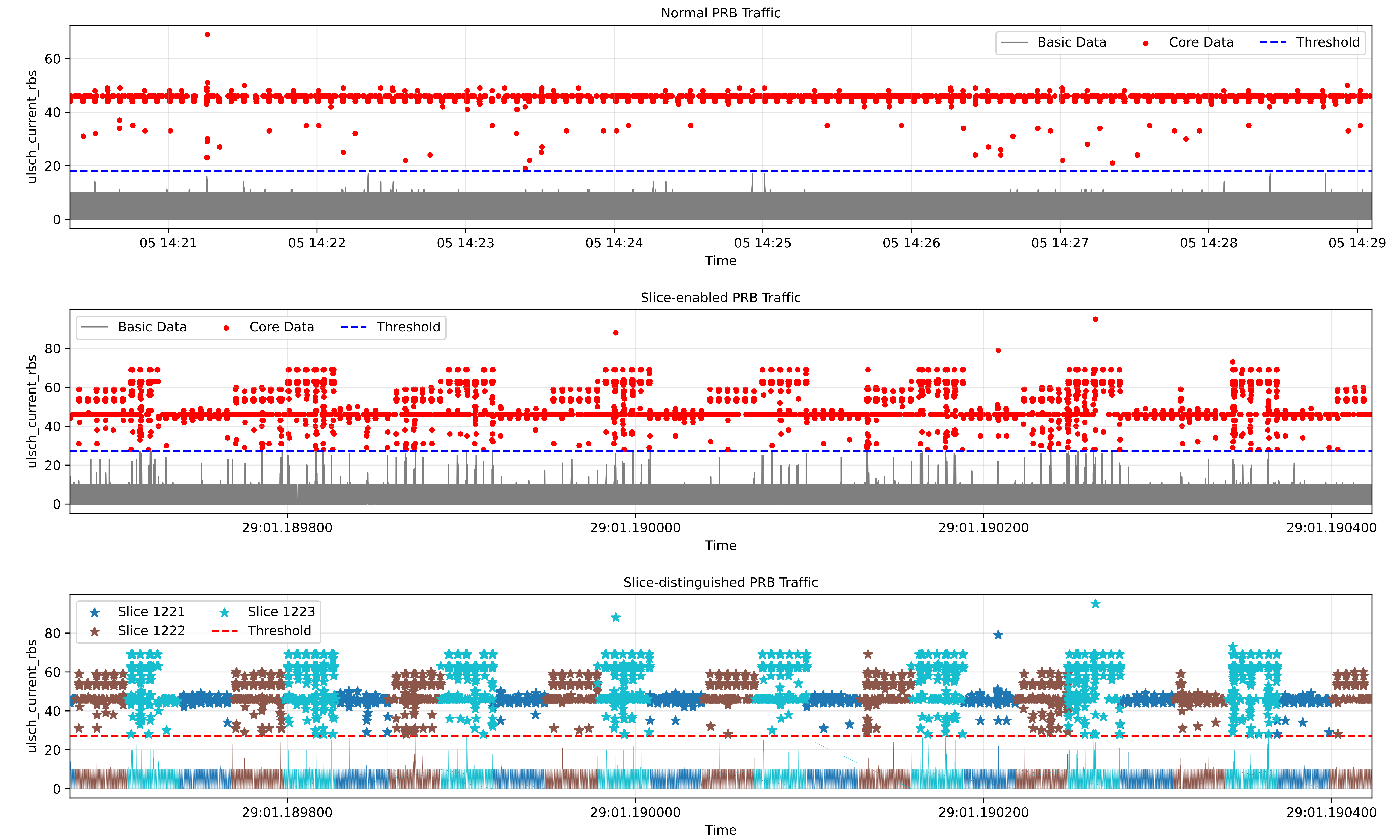}
    \caption{Temporal analysis of PRB allocation patterns across three implementation scenarios: normal traffic (top), slice-enabled traffic (middle), and slice-distinguished traffic (bottom). The x-axis represents timestamps in microseconds, while the y-axis denotes current resource blocks allocated to the UE. The visualization demonstrates the effectiveness of the slicing architecture in providing differentiated resource allocation.}
    \label{fig:Slice_RB}
\end{figure*}
\begin{figure*}[t]
    \centering
    \includegraphics[width=1\textwidth]{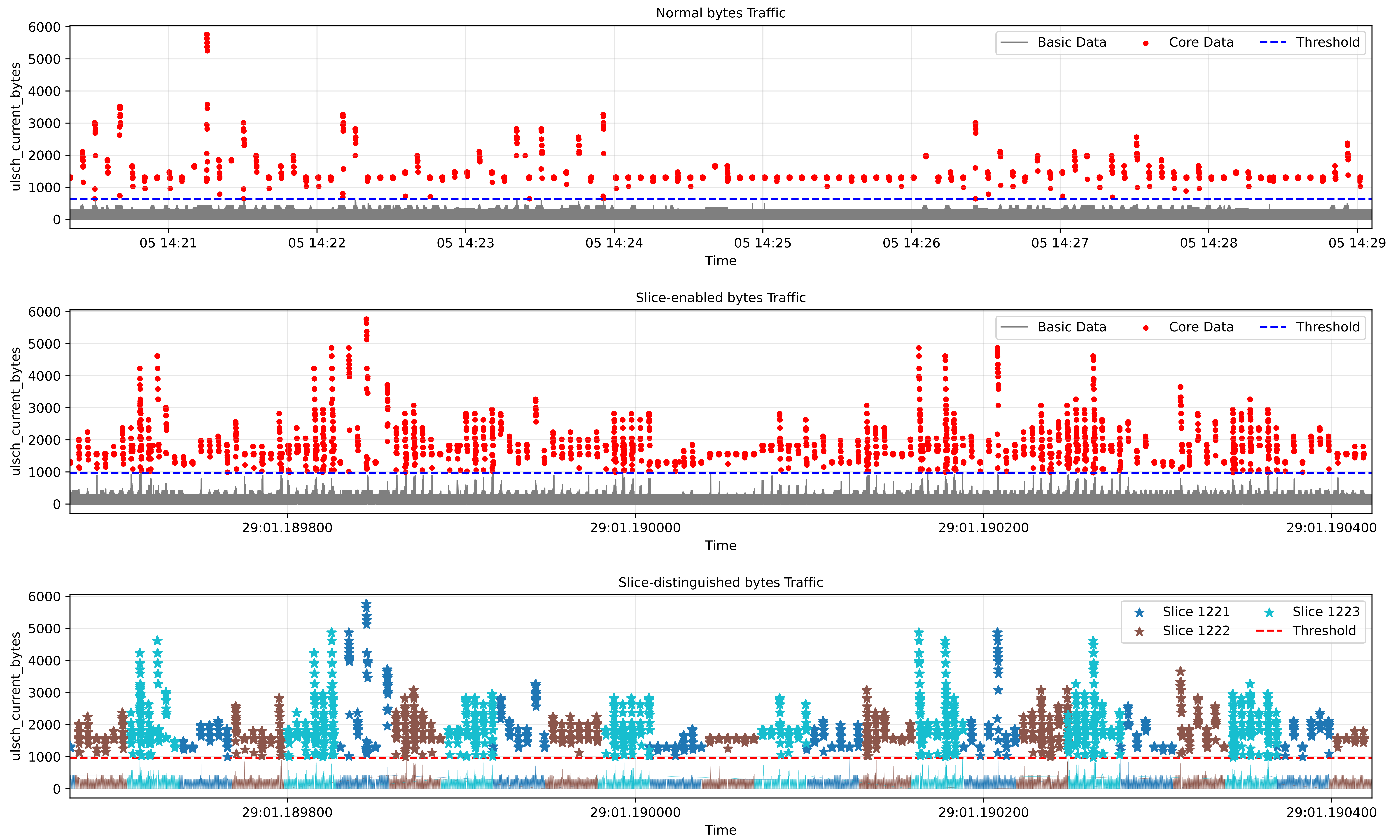}
    \caption{Temporal analysis of actual byte transmission across three scenarios: normal traffic (top), slice-enabled traffic (middle), and slice-distinguished traffic (bottom). Despite controlled PRB allocation, actual byte transmission shows high variance, particularly pronounced in LLM traffic patterns.}
    \label{fig:Slice_bytes}
\end{figure*}
Our measurements, visualized in Figures~\ref{fig:Slice_RB} and~\ref{fig:Slice_bytes}, reveal the practical implications of this unpredictability. Despite controlled Physical Resource Block (PRB) allocation following defined slice policies, actual byte transmission shows high variance and non-linear relationships with allocated resources. This non-linearity is particularly pronounced in LLM traffic compared to conventional data flows, as the stochastic generation process creates transmission bursts that poorly align with periodic resource scheduling.

The unpredictability extends beyond simple variance. Token streams exhibit complex temporal correlations as the model generates semantically coherent sequences, creating traffic patterns that defy traditional Markovian models. Resource allocation strategies based on average behavior or peak provisioning both prove inadequate: average-based allocation suffers from frequent deadline violations during generation bursts, while peak provisioning wastes resources during computation-bound phases. Effective LLM service delivery requires fine-grained monitoring and adaptive resource management that responds to real-time generation patterns rather than statistical traffic models.

These insights reveal why LLM services demand new wireless designs: bidirectional asymmetry requires direction-aware scheduling; computation-communication coupling needs dynamic slice optimization; and token stream unpredictability calls for adaptive resource management. Unlike traditional DNN applications, these characteristics necessitate rethinking wireless architectures. The following section presents WiLLM, our platform addressing these challenges through targeted design principles and implementation choices.

\section{WiLLM System Architecture}
\label{sec:architecture}
The insights from Section~\ref{sec:characteristics} drive WiLLM's architectural decisions. This section presents our design philosophy, system overview, and key innovations that collectively enable efficient LLM service delivery over wireless networks while democratizing access for researchers without wireless expertise.

\subsection{Design Philosophy}
\label{subsec:philosophy}
WiLLM's design philosophy centers on making LLM services accessible to researchers while addressing their unique technical challenges. Based on our empirical insights and the limitations of existing systems, we establish three core principles that guide our architectural decisions.

\textbf{Principle 1: Universal Accessibility.} Unlike existing testbeds that require specialized devices with native slicing support~\cite{saboorian2017network}, WiLLM must enable any standard device to leverage network slicing capabilities for fine-grained resource allocation. This principle drives our universal UE compatibility design through application-layer tunneling, allowing commodity devices to access slice-specific resources that guarantee LLM service requirements. By bridging the gap between slice-unaware devices and slice-enabled networks, we empower standard smartphones and laptops to benefit from differentiated QoS provisioning, ensuring predictable latency and bandwidth allocation crucial for interactive LLM applications. This democratizes advanced network resource management beyond specialized research equipment.

\textbf{Principle 2: Dynamic Adaptability.} The complex interplay between LLM computational variability and network conditions demands systems that can adapt in real-time. Static configurations fail to address the migrating bottlenecks and changing resource requirements inherent in LLM services. This principle motivates our dynamic slice compatibility mechanisms, multi-UE multi-slice scheduling capabilities, and dual-mode resource allocation, enabling the system to respond to varying workload characteristics without manual reconfiguration.

\textbf{Principle 3: Cross-Layer Observability.} The tight coupling between application-layer token generation, network-layer transmission, and computation-layer inference requires visibility and control across all system layers. Traditional layered architectures that hide internal state prevent the coordinated optimization necessary for LLM services. Real-time adaptation to LLM's dynamic behavior requires understanding not just what happens at each layer, but how events cascade across layers to impact end-to-end performance. This principle drives our comprehensive cross-layer API framework and monitoring infrastructure, exposing system state while maintaining appropriate abstractions for non-experts.

These principles balance technical requirements from our measurements with practical research needs. Rather than prescribing specific algorithms, WiLLM provides an open platform for deploying and evaluating arbitrary scheduling strategies, removing barriers that have limited innovation at the intersection of wireless networks and LLM services.

\subsection{Key System Innovations}
\label{subsec:overview}
\subsubsection{Core Network GPU Deployment}
\begin{figure*}[!htbp]
    \centering
    \includegraphics[width=1\textwidth]{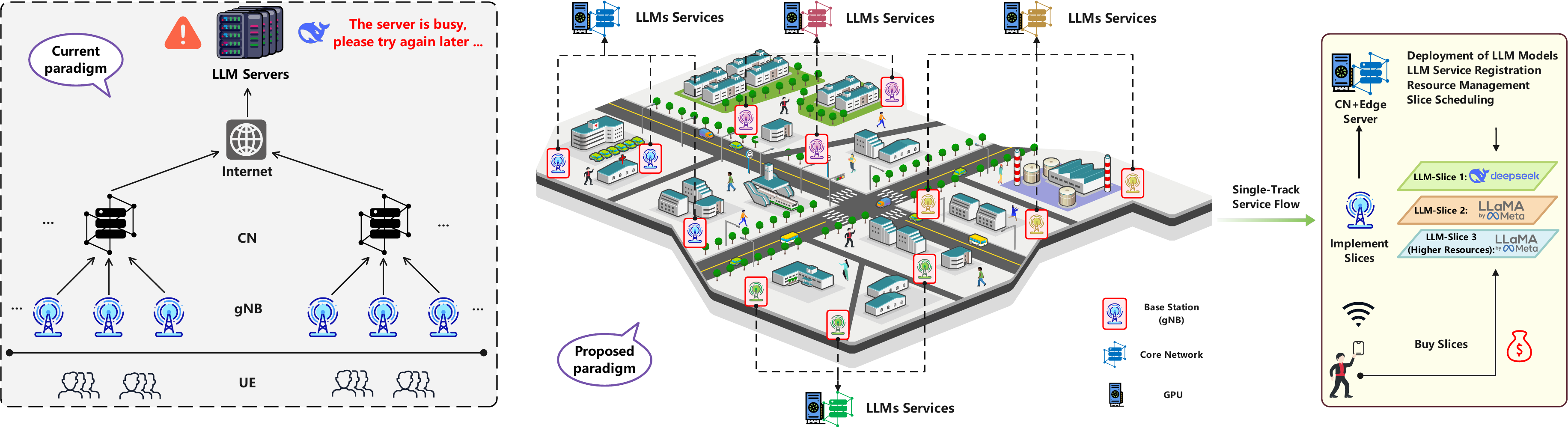}
    \caption{Left: Conventional cloud-based LLM service deployment architecture with internet connectivity. Right: Proposed core network LLM service deployment architecture with GPU resources at network convergence points.}
    \label{fig:core_network_deployment}
\end{figure*}
WiLLM fundamentally reimagines LLM service deployment by positioning GPU resources within the core network rather than at edge nodes or remote clouds. As illustrated in Figure~\ref{fig:core_network_deployment}, this architectural choice addresses the unique challenges of LLM services while leveraging existing telecommunications infrastructure.

Traditional cloud deployments (left side of Figure~\ref{fig:core_network_deployment}) require traffic to traverse the public internet, introducing variable latency, unpredictable congestion, and loss of operator control over service quality~\cite{chen2024netgpt}. Edge deployments attempt to reduce latency but face severe limitations: insufficient computational capacity for large language models, thermal constraints in compact edge nodes, and coordination challenges across distributed edge sites~\cite{he2024large, cai2024edge, nelson2023deploy, tang2023tinychat}.

Our core network deployment (right side of Figure~\ref{fig:core_network_deployment}) positions GPU clusters at the convergence point between high-capacity backhaul networks and the radio access network. This location offers unique advantages for LLM services. First, it provides access to substantial computational resources through dedicated GPU infrastructure connected via high-speed interconnects. Second, it maintains low latency by avoiding internet traversal while serving multiple base stations from a centralized location. Third, it enables coordinated resource management as both communication and computation resources fall under unified administrative control.

This deployment paradigm particularly benefits from the bidirectional asymmetry of LLM services. During uplink-heavy inference, the proximity to radio networks minimizes data transfer delays while powerful GPUs handle computational bottlenecks. For downlink-heavy generation, the high-bandwidth backhaul connections efficiently distribute generated content. The centralized nature also facilitates resource sharing across multiple concurrent LLM sessions, improving utilization compared to distributed edge deployments.

\subsubsection{Tree-Branch-Fruit Slicing Architecture}
\begin{figure}[!htbp]
    \centering
    \includegraphics[width=0.48\textwidth]{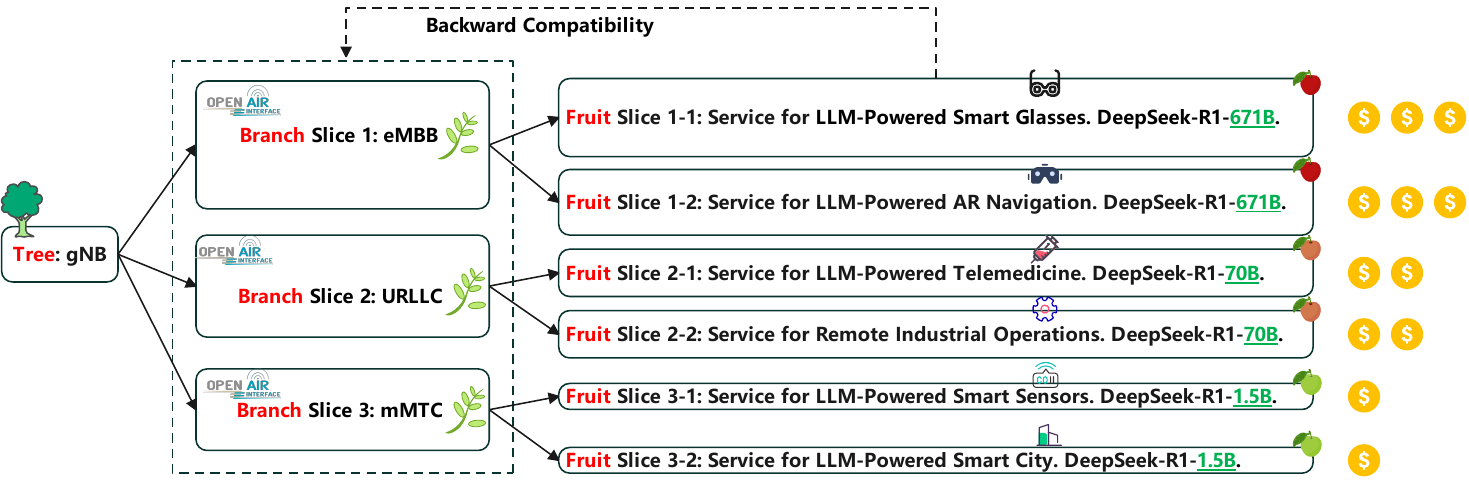}
    \caption{Tree-Branch-Fruit network slicing architecture with hierarchical service organization. The tree represents base infrastructure, branches correspond to traditional service categories, and fruits denote specialized LLM services with differentiated pricing models.}
    \label{fig:slicing_architecture}
\end{figure}

Building upon core network GPU deployment, WiLLM introduces the Tree-Branch-Fruit slicing architecture that extends traditional network slicing for LLM service differentiation. As depicted in Figure~\ref{fig:slicing_architecture}, this three-tier hierarchy enables flexible service provisioning while maintaining compatibility with existing 5G slicing mechanisms.

The \textbf{Tree} layer represents the foundational radio access network infrastructure shared across all services. This includes the physical base stations (gNBs), spectrum resources, and basic connectivity functions. The tree provides common entry points for all traffic while implementing isolation mechanisms to prevent interference between different service branches.

The \textbf{Branch} layer corresponds to traditional 5G network slices: enhanced Mobile Broadband (eMBB), Ultra-Reliable Low-Latency Communication (URLLC), and massive Machine Type Communication (mMTC). These branches maintain standard 3GPP-defined behaviors, ensuring backward compatibility with existing devices and services. Each branch implements specific quality-of-service parameters, resource allocation policies, and isolation guarantees as specified in 5G standards.

The \textbf{Fruit} layer introduces LLM-specific service slices that extend from appropriate branches based on their primary characteristics. Each fruit slice is configured for particular LLM applications with tailored resource allocations, model selections, and pricing tiers. For instance, a real-time translation service might extend from the URLLC branch with a lightweight model ensuring low latency, while a creative writing assistant could branch from eMBB with a larger model prioritizing quality over speed.

This hierarchical architecture enables several key capabilities. Service differentiation allows operators to offer multiple LLM service tiers with varying performance guarantees and pricing models, as indicated by the monetary symbols in Figure~\ref{fig:slicing_architecture}. Dynamic fruit creation permits rapid deployment of new LLM services without modifying core network functions. Resource isolation ensures that computational or communication demands from one LLM service do not impact others. Most importantly, the architecture maintains a clear evolutionary path where existing networks can progressively add LLM capabilities through fruit slice deployment.

\label{subsec:innovations}
\begin{figure*}[!htbp]
    \centering
    \includegraphics[width=0.98\textwidth]{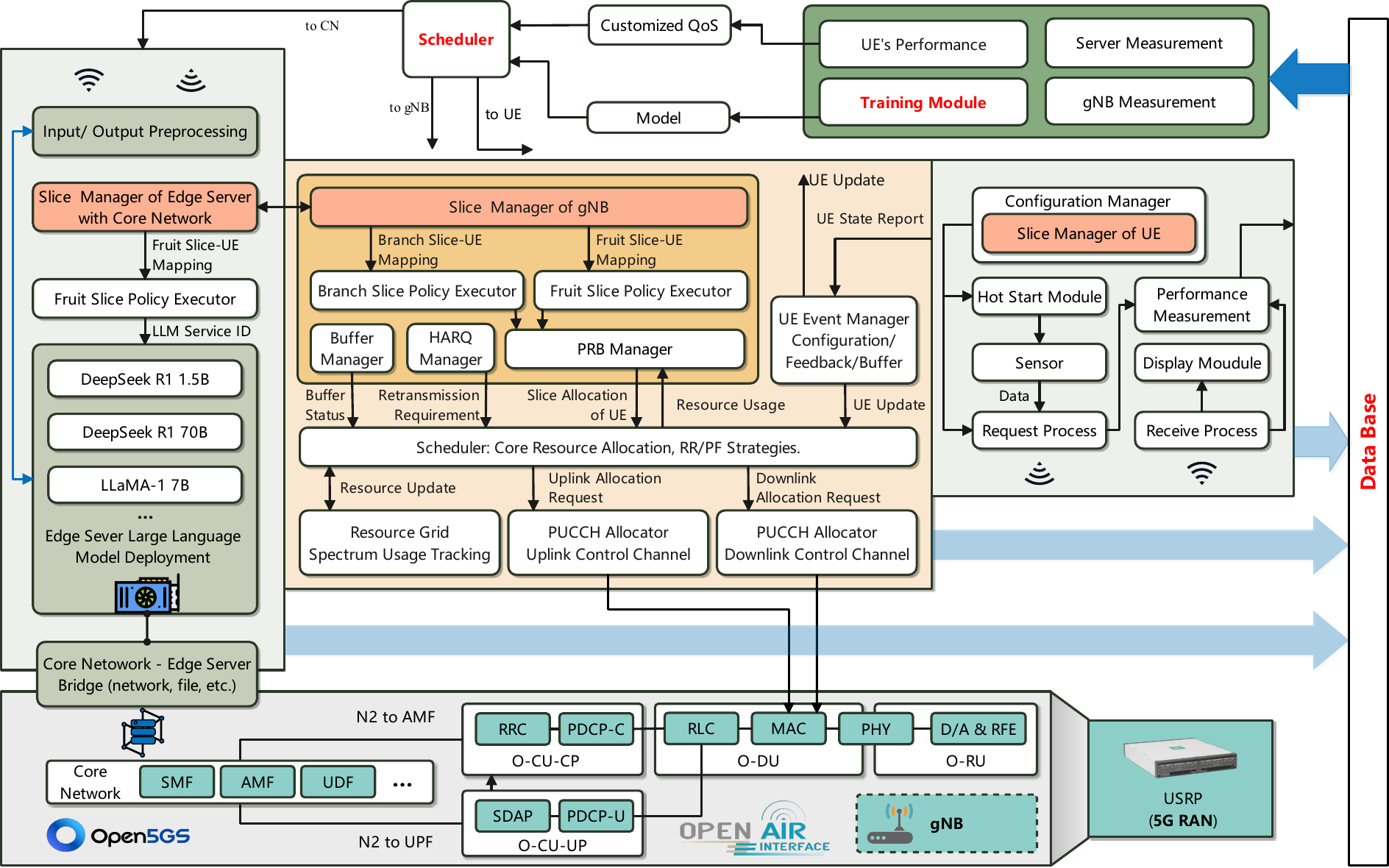}
    \caption{Comprehensive system architecture of WiLLM with UE, gNB, CN, and LLM Server Components. The diagram illustrates data flows, control pathways, and the integration of LLM-specific components within the telecommunications infrastructure.}
    \label{fig:system_architecture}
\end{figure*}

WiLLM implements five key innovations that collectively address the unique challenges of LLM service delivery while maintaining accessibility for researchers, as illustrated in the comprehensive system architecture shown in Figure~\ref{fig:system_architecture}.

\subsubsection{Dynamic Slice Compatibility}

Traditional network slicing assumes static service characteristics~\cite{andrade2025study}, but LLM services exhibit highly variable resource requirements based on model selection, generation parameters, and user interactions. WiLLM's dynamic slice compatibility extends conventional slicing with runtime adaptation capabilities.

As shown in Figure~\ref{fig:system_architecture}, the Slice Manager components in both gNB and UE subsystems implement bidirectional communication through UE Update and UE State Report pathways. This enables real-time slice reconfiguration based on observed performance metrics. The Branch Slice Policy Executor and Fruit Slice Policy Executor translate high-level service requirements into concrete resource allocation decisions, adapting to changing computational loads and network conditions.

The dynamic compatibility specifically addresses LLM's variable inference times and unpredictable token generation patterns. When the system detects a shift from computation-bound to communication-bound operation, slice parameters automatically adjust to maintain service quality. This adaptation occurs transparently, requiring no manual intervention from researchers.

\subsubsection{Universal UE Compatibility}

A critical barrier in existing platforms is the requirement for specialized UE equipment with native slicing support. WiLLM eliminates this barrier through application-layer tunneling mechanisms that enable any standard device to access LLM services.

The Slice Manager of UE component in Figure~\ref{fig:system_architecture} implements a middleware layer that intercepts application traffic and encapsulates it with appropriate slice identifiers. This approach bypasses the limitations of current UE implementations, which often lack complete slicing support. The Configuration Manager and Hot Start Module work together to establish slice contexts without requiring protocol-level modifications.

This innovation is crucial for democratizing access to LLM service research. Researchers can use commercial smartphones, laptops, or IoT devices without specialized firmware or hardware modifications. The application-layer approach also enables rapid prototyping of new service types without waiting for standardization or device updates.

\subsubsection{Multi-UE Multi-Slice Scheduling}
LLM services create complex scheduling challenges when multiple users with different service requirements share network resources. WiLLM implements sophisticated scheduling mechanisms that coordinate resources across multiple UEs and slice types simultaneously.

The Scheduler module at the center of Figure~\ref{fig:system_architecture} receives inputs from multiple sources: Buffer Status from the Buffer Manager, Retransmission Requirements from the HARQ Manager, and Slice Allocation policies from the mapping components. This multi-dimensional optimization problem is solved through a hierarchical approach: first allocating resources across slices based on their aggregate requirements, then scheduling individual UEs within each slice based on service-specific metrics.

The scheduling algorithm specifically accounts for LLM characteristics. Token generation bursts receive priority handling to minimize user-perceived latency, while background model updates are scheduled during resource valleys. The system maintains fairness across users while respecting the differentiated service levels promised by different fruit slices.

\subsubsection{Dual-Mode Resource Scheduling}

To balance implementation efficiency with optimization sophistication, WiLLM implements complementary embedded and separated scheduling modes, each suited to different operational scenarios.

In \textbf{embedded mode}, slice-aware resource allocation integrates directly within the main scheduling pipeline. This provides low-latency adaptation suitable for real-time LLM interactions. The scheduler modifies its decisions based on slice policies without external intervention, enabling microsecond-scale adaptations to changing conditions.

In \textbf{separated mode}, an external optimization engine analyzes longer-term patterns and computes globally optimal resource allocations. This engine interfaces through the Resource Update pathway, enabling sophisticated algorithms including machine learning-based prediction and multi-objective optimization. The separated mode excels at identifying persistent patterns in LLM usage and pre-allocating resources accordingly.

The dual-mode design allows researchers to experiment with different scheduling strategies without modifying core system components. Simple policies can be implemented in embedded mode for immediate testing, while complex algorithms can be developed in separated mode with full access to historical data and predictive models.

\subsubsection{Cross-Layer API Framework}
To enable effective research without requiring wireless expertise, WiLLM provides comprehensive APIs that abstract network complexity while exposing necessary control and monitoring capabilities. As shown in the online Figure~\href{https://openwillm.github.io}{16}, the API architecture implements a three-tier structure.

The \textbf{User Management API} handles service registration, authentication, and preference management. Researchers interact with familiar application-level concepts like "model selection" and "latency requirements" rather than radio network parameters. The API automatically translates these high-level specifications into appropriate network configurations.

The \textbf{System Management API} provides slice lifecycle management and service orchestration. Through RESTful interfaces, applications can discover available services, request specific configurations, and monitor performance metrics. The API includes WebSocket endpoints for real-time monitoring of token generation rates, latency breakdowns, and resource utilization.

The \textbf{Resource Management API} offers fine-grained control for advanced users who need to optimize specific aspects of service delivery. This includes interfaces for custom scheduling policies, resource reservation, and predictive allocation based on application hints. Importantly, this API remains optional—basic LLM services function without any low-level configuration.

The cross-layer design ensures information flows efficiently between system components. Token generation patterns from the application layer inform network scheduling decisions, while network congestion indicators trigger application-layer adaptation. This bidirectional information flow, illustrated by the various feedback pathways in Figure~\ref{fig:system_architecture}, enables holistic optimization impossible in traditional layered architectures.

These five innovations work synergistically to create a system that is both powerful enough to handle LLM's unique challenges and accessible enough for researchers without wireless expertise. The following section details how we realized these concepts through concrete implementation on the OpenAirInterface platform.
\section{Implementation}
\label{sec:implementation}

Transforming WiLLM's architectural vision into a functional system required substantial engineering effort. This section details our implementation approach, focusing on how we extended OpenAirInterface to support LLM services while maintaining compatibility and usability.

\subsection{Building on OpenAirInterface}
\label{subsec:oai_extension}

We selected OpenAirInterface (OAI) as our implementation foundation for several compelling reasons. First, OAI provides a complete, modular implementation of 4G/5G protocol stacks from physical layer to core network, enabling end-to-end system modifications. Second, its open-source nature allows deep modifications necessary for LLM-specific optimizations. Third, OAI's active community and extensive documentation lower the barrier for researchers to understand and extend our work. However, implementing WiLLM required substantial extensions beyond OAI's baseline capabilities. Our modifications span three critical areas:

\textbf{Slice Management Extensions.} While OAI includes basic network slicing support, it lacks the dynamic adaptation required for LLM services. We extended the Radio Resource Control (RRC) and Service Data Adaptation Protocol (SDAP) layers to support runtime slice reconfiguration. The modifications enable slice parameters to be adjusted based on real-time performance metrics without disrupting ongoing sessions. We implemented new signaling procedures that propagate slice updates from the core network through the RAN to individual UEs, maintaining consistency across the system.

\textbf{Scheduler Enhancements.} The OAI scheduler originally implements proportional fair and round-robin algorithms optimized for traditional traffic patterns. We augmented it with LLM-aware scheduling logic that recognizes token generation patterns and adapts resource allocation accordingly. Our modifications include: (1) traffic classification modules that identify LLM flows based on packet patterns, (2) dual-mode scheduling support with hooks for external optimization engines, and (3) direction-aware resource allocation that treats uplink and downlink asymmetrically based on identified bottlenecks.

\textbf{Cross-Layer Information Flow.} Traditional OAI maintains strict layer separation, preventing the coordinated optimization necessary for LLM services. We implemented information conduits that enable controlled cross-layer communication while preserving architectural modularity. New interfaces expose internal state from MAC, RLC, and PDCP layers to higher-level orchestration components. Similarly, application-layer hints about token generation patterns flow down to influence scheduling decisions. These modifications required careful design to avoid violating layer abstractions while enabling necessary information exchange.

\subsection{Engineering Challenges and Solutions}
\label{subsec:challenges}

Implementing WiLLM's architectural innovations presented several significant engineering challenges that required novel solutions based on the unique characteristics of LLM services identified in Section~\ref{sec:characteristics}.

\textbf{Challenge 1: Universal UE Compatibility Without Protocol Modifications.} A critical barrier in existing platforms is the requirement for specialized UE equipment with native slicing support. Current community implementations of network slicing remain incomplete and inconsistently supported across device ecosystems, effectively excluding standard commercial devices from accessing advanced network features.

\textit{Why existing systems fail:} Current platforms like OAI and srsRAN hard-code slice support at the protocol level, requiring modifications to UE protocol stacks for slice access. This approach excludes the vast majority of commercial devices and creates an impractical barrier for researchers who lack the capability to modify device firmware or develop custom UE implementations.

\textit{Solution:} We developed a middleware-based slice access mechanism through application-layer tunneling techniques. The Slice Manager of UE component implements a compatibility layer that encapsulates LLM service traffic within standard application protocols, enabling infrastructure compatibility while maintaining architectural autonomy. This approach provides telecommunications operators with implementation flexibility independent of underlying radio network slice capabilities, allowing fruit slices to evolve asynchronously from hardware upgrade cycles or standardization timelines.

\textbf{Challenge 2: Dynamic Slice Reconfiguration for Variable LLM Workloads.} LLM services exhibit highly variable resource requirements based on model selection, generation parameters, and user interactions, as demonstrated in our empirical analysis. Traditional network slicing assumes static service characteristics, creating a fundamental mismatch with LLM's dynamic nature.

\textit{Why existing systems fail:} Traditional slicing implementations in platforms like FlexRIC and Open5GS follow 3GPP specifications that assume static service profiles. Once configured, slice parameters remain fixed throughout the session lifetime. This rigid approach cannot accommodate LLM's dynamic resource requirements that vary by orders of magnitude within a single session.

\textit{Solution:} We extended conventional slicing with runtime adaptation capabilities through bidirectional communication pathways (UE Update and UE State Report) that enable real-time slice reconfiguration based on observed performance metrics. The Branch Slice Policy Executor and Fruit Slice Policy Executor translate high-level service requirements into concrete resource allocation decisions, adapting to changing computational loads and network conditions without disrupting ongoing sessions.

\textbf{Challenge 3: Cross-Layer Coordination for Computation - Communication Coupling.} Our measurements revealed tight coupling between LLM computational variability and network performance, where slice configuration changes cause dynamic bottleneck migration between computation and communication. Traditional layered architectures prevent the coordinated optimization necessary for managing this coupling.

\textit{Solution:} We implemented a comprehensive cross-layer API framework that facilitates vertical integration between application requirements and network-level resource allocation. The system includes performance feedback loops, service-specific optimization pathways that enable LLM Service ID information to influence network decisions, and resource utilization monitoring that provides visibility into consumption patterns across system layers while maintaining appropriate abstractions for non-experts.

\textbf{Challenge 4: Multi-UE Multi-Slice Scheduling Complexity.} LLM services create complex scheduling challenges when multiple users with different service requirements share network resources. The variable inference times and unpredictable token generation patterns identified in our analysis significantly complicate resource allocation compared to traditional traffic.

\textit{Solution:} We implemented a hierarchical scheduling mechanism that operates through a two-phase process: a global allocation phase distributes available physical resource blocks across active slices according to their relative priorities and resource guarantees, followed by an intra-slice scheduling phase that allocates resources to individual UEs within each slice based on service-specific metrics. This approach ensures fair and efficient resource utilization while maintaining service isolation.

\textbf{Challenge 5: Temporal Synchronization Across Distributed Components.} Correlating LLM service events across UE, RAN, and CN components requires precise time synchronization, particularly challenging given the variable processing delays inherent in LLM inference and the need for microsecond-level accuracy in performance analysis.

\textit{Solution:} We implemented NTP-based distributed synchronization across all system components, calibrating devices through a common server to achieve microsecond-level sampling precision. By recording timestamps at both UE and server endpoints and implementing latency compensation algorithms, we maintained synchronization errors within ±1.0 milliseconds. This precision enables the accurate cross-layer correlation analysis that forms the foundation of our empirical insights.

These engineering solutions demonstrate that WiLLM's architectural innovations can be realized through practical implementations that address the unique challenges of LLM service delivery while maintaining system stability and research accessibility.
\section{WiLLM in Action: Deployment, Dataset, and Case Study}
\label{sec:evaluation}

This section demonstrates WiLLM's practical capabilities through comprehensive deployment, data collection, and real-world application development. Unlike conventional wireless testbeds that focus on protocol compliance, WiLLM enables researchers to study and optimize LLM services across realistic wireless environments.

\subsection{Real-World Deployment}
\label{subsec:deployment}

WiLLM's operational deployment demonstrates the system's capability to deliver end-to-end LLM services while providing a robust experimentation platform for researchers.


\textbf{System Performance and Capabilities:} As illustrated in our online appendix Figure~\href{https://openwillm.github.io}{14}, our deployed testbed achieves full end-to-end LLM service delivery with commercial devices. The system successfully handles concurrent multi-user sessions with differentiated service levels through the Tree-Branch-Fruit slicing architecture. The deployment maintains stable operation across diverse LLM workloads, from lightweight text generation to compute-intensive multimodal processing, validating the practical viability of core network GPU deployment for production environments.

\textbf{Research Platform Features:} The testbed provides researchers with unprecedented experimental control through its integrated monitoring and adaptation capabilities. Real-time performance visualization enables immediate observation of cross-layer interactions, while programmable APIs allow dynamic reconfiguration without service interruption. Researchers can experiment with various LLM models (LLaVA, Llama 3.2) and network configurations through simple parameter adjustments, without requiring telecommunications expertise. The platform's stability and ease of use have enabled extended experimental campaigns lasting multiple days without manual intervention.

\textbf{Accessibility and Adoption:} A key achievement is enabling AI researchers to experiment with communication resource allocation without deep protocol stack expertise. Through intuitive Python APIs and web interfaces, researchers can specify scheduling policies, adjust slice parameters, and control bandwidth allocation—making wireless resource management accessible to those focused on LLM optimization. The platform abstracts complex protocol details while exposing meaningful control knobs: AI researchers can explore how different QoS configurations, priority schemes, and resource partitioning strategies impact their models' performance. Standard devices connect seamlessly through our application-layer approach, allowing experiments on commodity hardware. This design bridges the gap between AI and networking communities, enabling AI researchers to actively engage with communication-layer optimizations rather than treating the network as a black box.

\subsection{Dataset}
\label{subsec:dataset}
The WiLLM dataset represents the first comprehensive empirical foundation specifically designed for wireless LLM research, providing synchronized multi-layer measurements essential for algorithm development and validation.

\textbf{Dataset Architecture and Coverage:} Our dataset comprises 1,649,996 synchronized records collected across 58 metrics spanning three system layers. The measurements cover four distinct operational scenarios with varying dynamics: static UE with static gNB (290,653 records), dynamic UE with static gNB (363,906 records), static UE with dynamic gNB (430,369 records), and dynamic UE with dynamic gNB (565,068 records). This systematic variation enables researchers to analyze system behavior under different mobility and resource management conditions.

\textbf{Multi-Layer Metric Synchronization:} The dataset captures 22 UE-layer metrics including request latency, response timing, image resolution parameters (320×240 to 640×480 pixels), response word count (50-200 words), and payload characteristics. At the gNB layer, 25 metrics document radio conditions (MCS, CQI, BLER, SNR), PRB allocation patterns, slice configurations, and scheduling decisions. The Core/Edge layer contributes 18 metrics measuring LLM inference timing, token generation rates, model selection (LLaVA, Llama 3.2), and GPU utilization patterns. Through NTP-based synchronization maintaining ±1.0ms precision, these metrics enable accurate correlation between user experience and system-level events.

\textbf{Dataset Uniqueness:} Unlike traditional wireless datasets focused on channel measurements or throughput statistics, WiLLM's dataset explicitly captures the interplay between token generation patterns, wireless resource allocation, and user-perceived performance. The synchronized multi-layer nature reveals complex interactions impossible to observe through isolated measurements. The temporal synchronization enables researchers to correlate LLM inference events with network resource allocation decisions, exposing the dynamic bottleneck migration between computation and communication that characterizes LLM services.

\textbf{Reproducibility and Standardization:} The dataset includes comprehensive documentation with detailed metric definitions (the online Table~\href{https://openwillm.github.io}{4}) and collection parameters, enabling proper utilization and result reproduction. By providing standardized measurement data, researchers can evaluate their algorithms against the same realistic conditions, ensuring fair comparison across different approaches. This standardization is crucial for advancing the field beyond isolated demonstrations toward systematic performance improvements.

\subsection{Case Study: Smart Glasses Application}
\label{subsec:casestudy}
To demonstrate WiLLM's practical utility for application development, we implemented a smart glasses system that validates our platform's ability to support real-world LLM applications while addressing human-computer interaction requirements.

\subsubsection{Application Architecture and Integration}
\begin{figure}[htbp]
\centering
\includegraphics[width=0.48\textwidth]{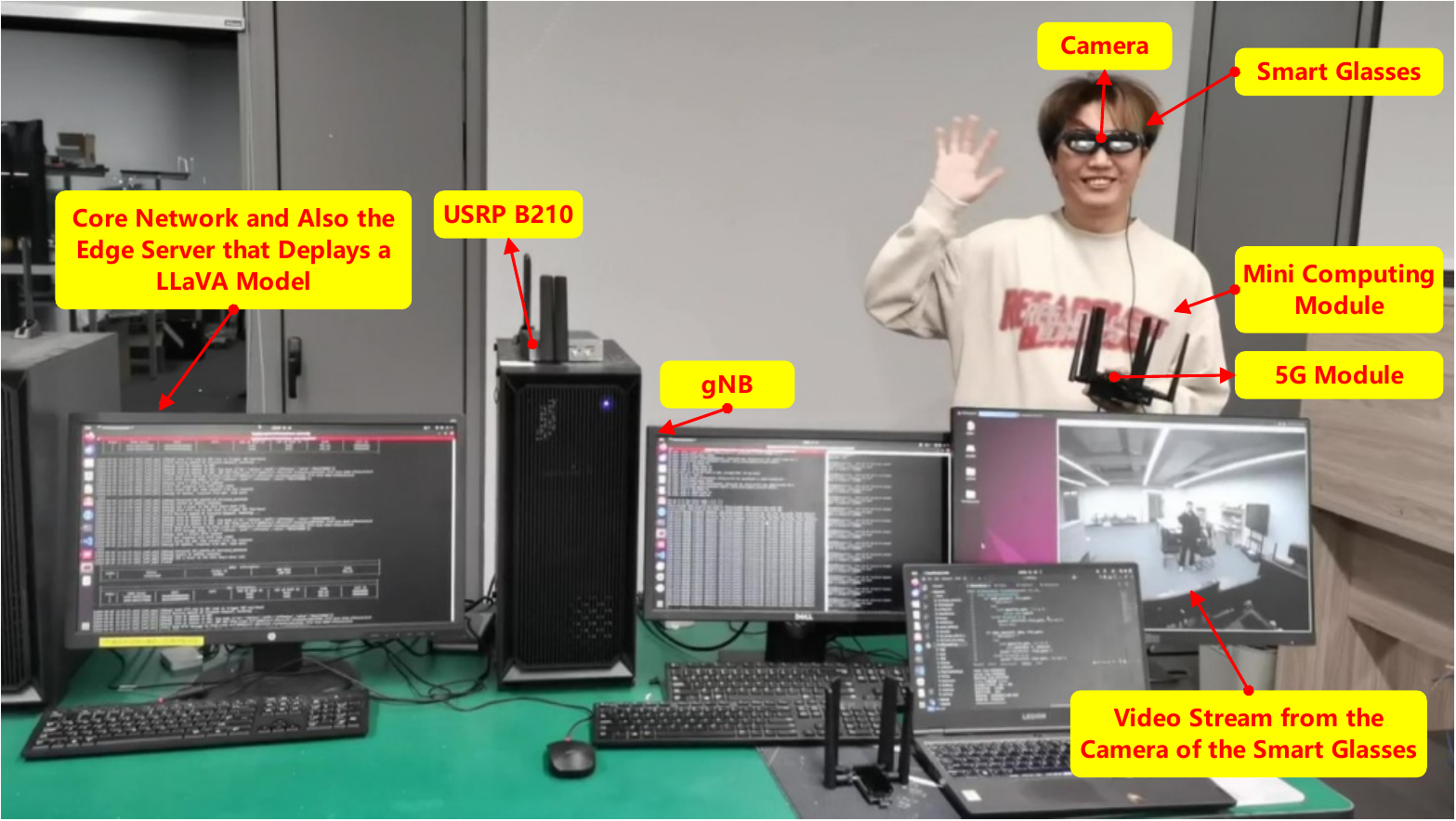}
\caption{Smart glasses integrated with WiLLM. The system consists of customized smart glasses paired with a compact processing unit housing the 5G module, demonstrating practical deployment of our Tree-Branch-Fruit architecture.}
\label{fig:SmartGlassesScenario}
\end{figure}

Our smart glasses implementation, illustrated in Figure~\ref{fig:SmartGlassesScenario}, represents a quintessential validation of WiLLM's design principles. The hardware comprises custom-built smart glasses with an integrated camera positioned at the glabella for natural field-of-view capture, paired with a compact processing unit containing a 5G module and minimal computational resources. This configuration validates our architectural principle that resource-constrained devices can access sophisticated LLM capabilities through core network deployment.

The system architecture follows WiLLM's design precisely: glasses function as intelligent sensors and display interfaces while computational processing occurs at GPU-equipped core network infrastructure. This separation validates our decision to position LLM resources at network convergence points rather than attempting resource-intensive edge deployment.

\begin{figure}[htbp]
\centering
\includegraphics[width=0.45\textwidth]{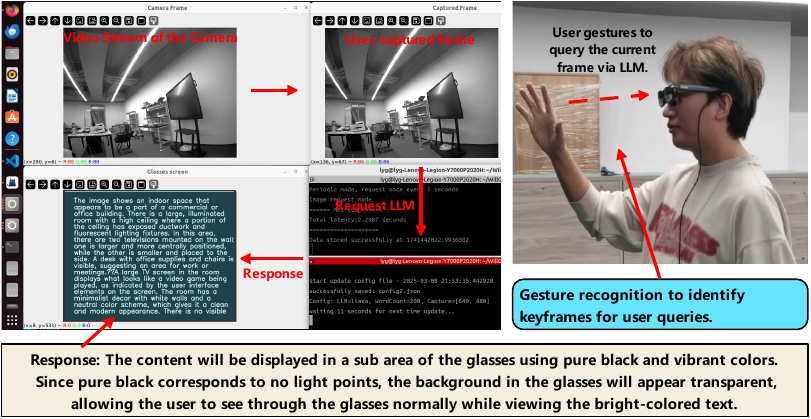}
\caption{Gesture-triggered interaction paradigm demonstrating seamless integration between physical actions and LLM service delivery through WiLLM's bidirectional communication capabilities.}
\label{fig:Gesture}
\end{figure}

The interaction paradigm, shown in Figure~\ref{fig:Gesture}, demonstrates practical implementation of our bidirectional communication architecture. Users trigger scene analysis through intuitive gestures—extending five fingers followed by grasping motion—which captures visual frames and initiates LLM queries. This workflow validates WiLLM's capability to handle the asymmetric traffic patterns identified in Section~\ref{sec:characteristics}: high-resolution images flow uplink for scene understanding while generated descriptions return downlink for display.

\subsubsection{Human-Computer Interaction Optimization}
\textbf{Performance Objectives:} Human-computer interaction research establishes that user satisfaction depends more on predictable response patterns than absolute minimal latency ~\cite{jacobs1997managing, nabiyouni2017relative}. This principle guided our optimization target: achieving stable 2-second response latency rather than pursuing minimal but variable delays. This objective balances perceptual continuity, technical feasibility, and system reliability within realistic network conditions.

\textbf{Optimization Methodologies:} We employed complementary approaches demonstrating WiLLM's versatility for algorithm development:

\textit{Offline Analysis:} Leveraging WiLLM's comprehensive data collection capabilities, we systematically evaluated six slice configurations across extended operational periods. Each configuration was tested with diverse visual scenes and varying network conditions to build statistical performance models. Analysis revealed that intermediate slice allocations provided optimal balance between resource efficiency and performance consistency.

\textit{Online Learning:} We implemented an Upper Confidence Bound (UCB) algorithm that continuously optimizes slice selection based on observed performance. The algorithm initially explores different configurations, gradually converging toward those consistently meeting latency targets.

\begin{figure}[t]
\centering
\includegraphics[width=0.48\textwidth]{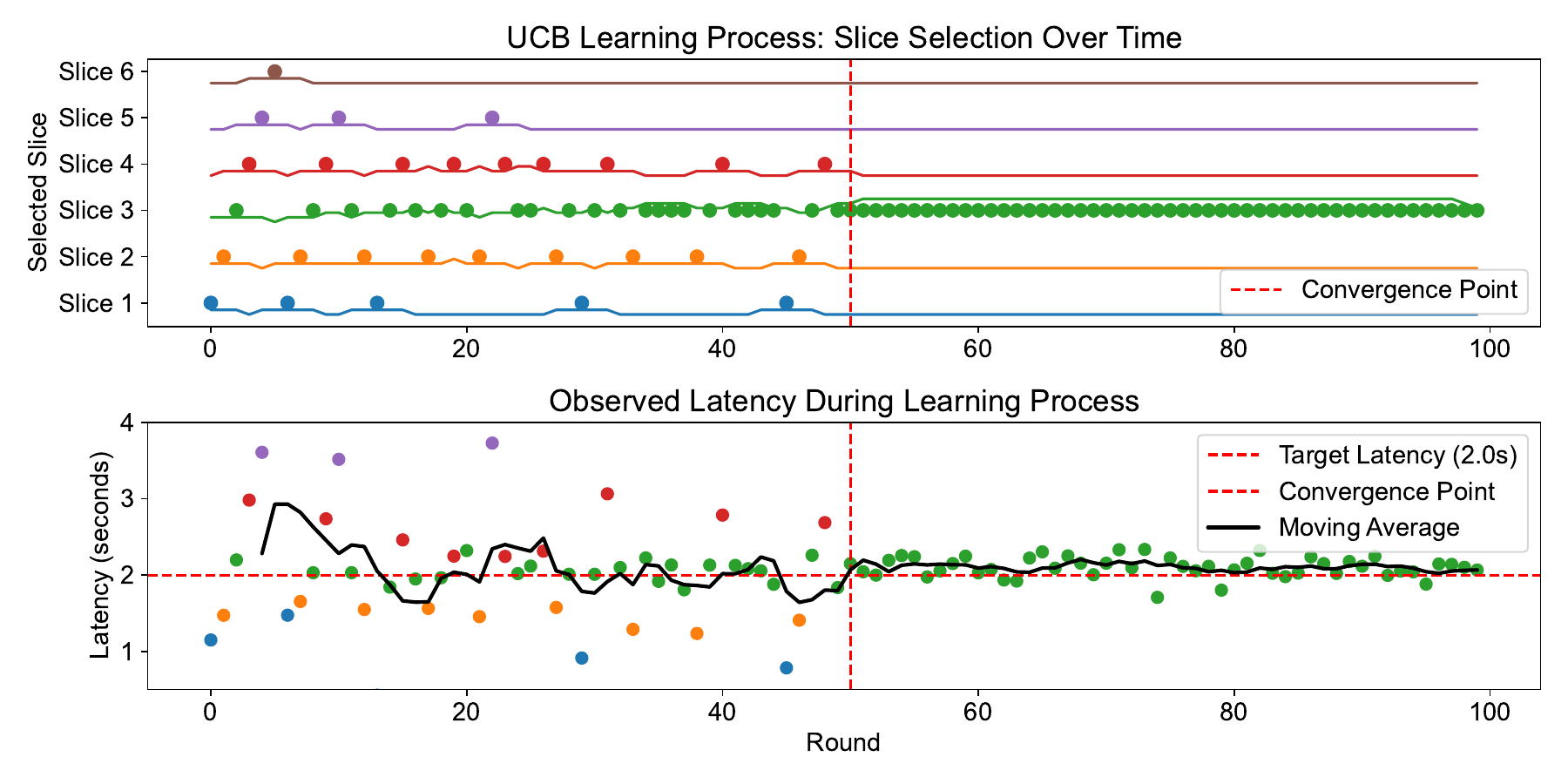}
\caption{Online learning progress for slice selection optimization, demonstrating WiLLM's support for adaptive algorithm development and validation.}
\label{fig:UCB}
\end{figure}

As shown in Figure~\ref{fig:UCB}, the online approach adapts to non-stationary conditions as network load varies or LLM model updates change computational requirements. This capability validates WiLLM's cross-layer APIs that provide real-time feedback enabling dynamic optimization.
\section{Related Works}
\label{sec:related}

\subsection{Wireless Systems for AI/ML Services}
\label{subsec:related_wireless}

Recent research has explored various approaches to supporting AI/ML workloads over wireless networks. Edge computing frameworks~\cite{chen2024netgpt, wu2020deep} position computational resources at network edges to reduce latency, but face fundamental limitations in supporting large language models due to power and thermal constraints. Cloud-edge collaborative architectures~\cite{dong2024creating} attempt to balance computation between edge and cloud, yet struggle with the tight coupling between LLM inference and communication that our measurements reveal.

Distributed inference systems~\cite{xue2024wdmoe, mitra2024distributed} partition LLM computation across multiple nodes. While these approaches address computational scalability, they introduce complex coordination challenges and remain largely theoretical without practical wireless implementations. In contrast, WiLLM provides a complete, operational system that researchers can immediately use for experimentation.

Network slicing for AI services has received attention~\cite{liu2024llm, dandoush2024large}, but existing work focuses on simulation-based evaluation rather than real implementations. These studies identify the potential of slicing for service differentiation but lack the dynamic adaptation mechanisms necessary for LLM's variable workloads. WiLLM's Tree-Branch-Fruit architecture extends beyond static slicing to provide runtime adaptation based on observed performance.
\vspace{-6pt}
\subsection{Open-Source Wireless Platforms}
\label{subsec:related_platforms}
Several open-source platforms provide wireless network implementations, each with distinct limitations for LLM service research. OpenAirInterface~\cite{nikaein2014openairinterface} offers comprehensive protocol implementations but requires deep wireless expertise and lacks LLM-specific features. srsRAN~\cite{gomez2016srslte} provides a lightweight alternative but similarly focuses on traditional communication services. Open5GS~\cite{barbosa2024open} implements core network functions without radio access capabilities, limiting end-to-end experimentation. Recent platforms like FlexRIC~\cite{schmidt2021flexric} and Janus~\cite{foukas2023taking} introduce programmability for network intelligence but remain focused on traditional metrics and lack the cross-layer visibility essential for LLM services. Critically, all existing platforms require specialized UE hardware or protocol modifications for advanced features like network slicing, creating barriers for AI researchers. WiLLM addresses these limitations through our universal compatibility approach and comprehensive APIs that abstract wireless complexity. By building on OAI's solid foundation while adding LLM-specific innovations, we provide a platform accessible to researchers without telecommunications expertise.

\subsection{LLM Deployment and Optimization}
\label{subsec:related_llm}

The LLM deployment landscape includes various frameworks optimizing inference performance~\cite{tensorrt_llm, llama_cpp}, but these focus solely on computational efficiency without considering wireless communication constraints. Container-based solutions like Ollama~\cite{ollama} simplify deployment but lack awareness of network conditions that significantly impact service quality in wireless environments.

Recent work on LLM serving systems~\cite{xu2025serving, zhang2025beyond} addresses challenges like long-context handling and request scheduling in datacenter environments. However, these solutions assume stable, high-bandwidth connectivity unavailable in wireless settings. WiLLM bridges this gap by providing network awared LLM serving that adapts to wireless channel variations and resource constraints.

This survey of existing work underscores the absence of integrated solutions for wireless LLM service delivery. The dichotomy between wireless infrastructure platforms and LLM deployment frameworks has created a research void that WiLLM fills through its unified architecture, empirical foundation, and accessible implementation. Our contributions demonstrate that bridging these previously separate domains enables new research opportunities and practical applications impossible with existing fragmented approaches.
\section{Conclusion}
\label{sec:conclusion}
We presented WiLLM, the first open-source platform for wireless LLM service delivery. We identified three characteristics of LLM services over wireless network: bidirectional asymmetry, computation-communication coupling, and token unpredictability. WiLLM addresses these through the new slicing technology, universal compatibility, and cross-layer APIs, enabling AI researchers to experiment with real cellular networks without wireless expertise. The smart glasses case study validated our approach with stable 2-second multimodal responses. By democratizing wireless LLM research, WiLLM transforms a specialized domain into an accessible field for advancing mobile AI services.

\clearpage
\bibliographystyle{ACM-Reference-Format}
\bibliography{ref}


\begin{thebibliography}{41}


\ifx \showCODEN    \undefined \def \showCODEN     #1{\unskip}     \fi
\ifx \showISBNx    \undefined \def \showISBNx     #1{\unskip}     \fi
\ifx \showISBNxiii \undefined \def \showISBNxiii  #1{\unskip}     \fi
\ifx \showISSN     \undefined \def \showISSN      #1{\unskip}     \fi
\ifx \showLCCN     \undefined \def \showLCCN      #1{\unskip}     \fi
\ifx \shownote     \undefined \def \shownote      #1{#1}          \fi
\ifx \showarticletitle \undefined \def \showarticletitle #1{#1}   \fi
\ifx \showURL      \undefined \def \showURL       {\relax}        \fi
\providecommand\bibfield[2]{#2}
\providecommand\bibinfo[2]{#2}
\providecommand\natexlab[1]{#1}
\providecommand\showeprint[2][]{arXiv:#2}

\bibitem[Alsmirat et~al\mbox{.}(2023)]%
        {alsmirat2023video}
\bibfield{author}{\bibinfo{person}{Mohammad Alsmirat}, \bibinfo{person}{Yousef Sharrab}, \bibinfo{person}{Monther Tarawneh}, \bibinfo{person}{Sana’a Al-shboul}, {and} \bibinfo{person}{Nabil Sarhan}.} \bibinfo{year}{2023}\natexlab{}.
\newblock \showarticletitle{Video coding deep learning-based modeling for long life video streaming over next network generation}.
\newblock \bibinfo{journal}{\emph{Cluster Computing}} \bibinfo{volume}{26}, \bibinfo{number}{2} (\bibinfo{year}{2023}), \bibinfo{pages}{1159--1167}.
\newblock


\bibitem[Andrade and Wickboldt(2025)]%
        {andrade2025study}
\bibfield{author}{\bibinfo{person}{Maiko Andrade} {and} \bibinfo{person}{Juliano~Araujo Wickboldt}.} \bibinfo{year}{2025}\natexlab{}.
\newblock \showarticletitle{A Study on 5G Network Slice Isolation Based on Native Cloud and Edge Computing Tools}.
\newblock \bibinfo{journal}{\emph{arXiv preprint arXiv:2502.02842}} (\bibinfo{year}{2025}).
\newblock


\bibitem[Barbosa et~al\mbox{.}(2024)]%
        {barbosa2024open}
\bibfield{author}{\bibinfo{person}{Maria Barbosa}, \bibinfo{person}{Marcelo Silva}, \bibinfo{person}{Ednelson Cavalcanti}, {and} \bibinfo{person}{Kelvin Dias}.} \bibinfo{year}{2024}\natexlab{}.
\newblock \showarticletitle{Open-Source 5G Core Platforms: A Low-Cost Solution and Performance Evaluation}.
\newblock \bibinfo{journal}{\emph{arXiv preprint arXiv:2412.21162}} (\bibinfo{year}{2024}).
\newblock


\bibitem[Bariah et~al\mbox{.}(2024)]%
        {bariah2024large}
\bibfield{author}{\bibinfo{person}{Lina Bariah}, \bibinfo{person}{Qiyang Zhao}, \bibinfo{person}{Hang Zou}, \bibinfo{person}{Yu Tian}, \bibinfo{person}{Faouzi Bader}, {and} \bibinfo{person}{Merouane Debbah}.} \bibinfo{year}{2024}\natexlab{}.
\newblock \showarticletitle{Large generative ai models for telecom: The next big thing?}
\newblock \bibinfo{journal}{\emph{IEEE Communications Magazine}} \bibinfo{volume}{62}, \bibinfo{number}{11} (\bibinfo{year}{2024}), \bibinfo{pages}{84--90}.
\newblock


\bibitem[Boateng et~al\mbox{.}(2024)]%
        {boateng2024survey}
\bibfield{author}{\bibinfo{person}{Gordon~Owusu Boateng}, \bibinfo{person}{Hani Sami}, \bibinfo{person}{Ahmed Alagha}, \bibinfo{person}{Hanae Elmekki}, \bibinfo{person}{Ahmad Hammoud}, \bibinfo{person}{Rabeb Mizouni}, \bibinfo{person}{Azzam Mourad}, \bibinfo{person}{Hadi Otrok}, \bibinfo{person}{Jamal Bentahar}, \bibinfo{person}{Sami Muhaidat}, {et~al\mbox{.}}} \bibinfo{year}{2024}\natexlab{}.
\newblock \showarticletitle{A Survey on Large Language Models for Communication, Network, and Service Management: Application Insights, Challenges, and Future Directions}.
\newblock \bibinfo{journal}{\emph{arXiv preprint arXiv:2412.19823}} (\bibinfo{year}{2024}).
\newblock


\bibitem[Cai et~al\mbox{.}(2024)]%
        {cai2024edge}
\bibfield{author}{\bibinfo{person}{Fenglong Cai}, \bibinfo{person}{Dong Yuan}, \bibinfo{person}{Zhe Yang}, {and} \bibinfo{person}{Lizhen Cui}.} \bibinfo{year}{2024}\natexlab{}.
\newblock \showarticletitle{Edge-llm: A collaborative framework for large language model serving in edge computing}. In \bibinfo{booktitle}{\emph{2024 IEEE International Conference on Web Services (ICWS)}}. IEEE, \bibinfo{pages}{799--809}.
\newblock


\bibitem[Chaoub and Elkotob(2025)]%
        {chaoub2025mobile}
\bibfield{author}{\bibinfo{person}{Abdelaali Chaoub} {and} \bibinfo{person}{Muslim Elkotob}.} \bibinfo{year}{2025}\natexlab{}.
\newblock \showarticletitle{Mobile Network-specialized Large Language Models for 6G: Architectures, Innovations, Challenges, and Future Trends}.
\newblock \bibinfo{journal}{\emph{arXiv preprint arXiv:2502.04933}} (\bibinfo{year}{2025}).
\newblock


\bibitem[Chen et~al\mbox{.}(2024)]%
        {chen2024netgpt}
\bibfield{author}{\bibinfo{person}{Yuxuan Chen}, \bibinfo{person}{Rongpeng Li}, \bibinfo{person}{Zhifeng Zhao}, \bibinfo{person}{Chenghui Peng}, \bibinfo{person}{Jianjun Wu}, \bibinfo{person}{Ekram Hossain}, {and} \bibinfo{person}{Honggang Zhang}.} \bibinfo{year}{2024}\natexlab{}.
\newblock \showarticletitle{NetGPT: An AI-native network architecture for provisioning beyond personalized generative services}.
\newblock \bibinfo{journal}{\emph{IEEE Network}} (\bibinfo{year}{2024}).
\newblock


\bibitem[Dandoush et~al\mbox{.}(2024)]%
        {dandoush2024large}
\bibfield{author}{\bibinfo{person}{Abdulhalim Dandoush}, \bibinfo{person}{Viswanath Kumarskandpriya}, \bibinfo{person}{Mueen Uddin}, {and} \bibinfo{person}{Usman Khalil}.} \bibinfo{year}{2024}\natexlab{}.
\newblock \showarticletitle{Large language models meet network slicing management and orchestration}.
\newblock \bibinfo{journal}{\emph{arXiv preprint arXiv:2403.13721}} (\bibinfo{year}{2024}).
\newblock


\bibitem[Dong et~al\mbox{.}(2024)]%
        {dong2024creating}
\bibfield{author}{\bibinfo{person}{Qifei Dong}, \bibinfo{person}{Xiangliang Chen}, {and} \bibinfo{person}{Mahadev Satyanarayanan}.} \bibinfo{year}{2024}\natexlab{}.
\newblock \showarticletitle{Creating edge ai from cloud-based llms}. In \bibinfo{booktitle}{\emph{Proceedings of the 25th International Workshop on Mobile Computing Systems and Applications}}. \bibinfo{pages}{8--13}.
\newblock


\bibitem[Flakowski et~al\mbox{.}(2023)]%
        {flakowski2023implementation}
\bibfield{author}{\bibinfo{person}{Wojciech Flakowski}, \bibinfo{person}{Maciej Krasicki}, {and} \bibinfo{person}{Rafa{\l} Krenz}.} \bibinfo{year}{2023}\natexlab{}.
\newblock \showarticletitle{Implementation of a 4g/5g base station using the srsran software and the usrp software radio module}.
\newblock \bibinfo{journal}{\emph{Journal of Telecommunications and Information Technology}} \bibinfo{number}{3} (\bibinfo{year}{2023}), \bibinfo{pages}{30--40}.
\newblock


\bibitem[Foukas et~al\mbox{.}(2023)]%
        {foukas2023taking}
\bibfield{author}{\bibinfo{person}{Xenofon Foukas}, \bibinfo{person}{Bozidar Radunovic}, \bibinfo{person}{Matthew Balkwill}, {and} \bibinfo{person}{Zhihua Lai}.} \bibinfo{year}{2023}\natexlab{}.
\newblock \showarticletitle{Taking 5G RAN analytics and control to a new level}. In \bibinfo{booktitle}{\emph{Proceedings of the 29th Annual International Conference on Mobile Computing and Networking}}. \bibinfo{pages}{1--16}.
\newblock


\bibitem[{ggml-org}(2025)]%
        {llama_cpp}
\bibfield{author}{\bibinfo{person}{{ggml-org}}.} \bibinfo{year}{2025}\natexlab{}.
\newblock \bibinfo{title}{{llama.cpp: LLM inference in C/C++}}.
\newblock
\urldef\tempurl%
\url{https://github.com/ggml-org/llama.cpp}
\showURL{%
\tempurl}
\newblock
\shownote{Accessed: 2025-02-27}.


\bibitem[Gomez-Miguelez et~al\mbox{.}(2016)]%
        {gomez2016srslte}
\bibfield{author}{\bibinfo{person}{Ismael Gomez-Miguelez}, \bibinfo{person}{Andres Garcia-Saavedra}, \bibinfo{person}{Paul~D. Sutton}, \bibinfo{person}{Pablo Serrano}, \bibinfo{person}{Cristina Cano}, {and} \bibinfo{person}{Douglas~J. Leith}.} \bibinfo{year}{2016}\natexlab{}.
\newblock \showarticletitle{srsLTE: An Open-Source Platform for LTE Evolution and Experimentation}. In \bibinfo{booktitle}{\emph{Proceedings of the Tenth ACM International Workshop on Wireless Network Testbeds, Experimental Evaluation, and Characterization (WiNTECH '16)}}. \bibinfo{publisher}{ACM}, \bibinfo{pages}{25--32}.
\newblock


\bibitem[He et~al\mbox{.}(2024)]%
        {he2024large}
\bibfield{author}{\bibinfo{person}{Ying He}, \bibinfo{person}{Jingcheng Fang}, \bibinfo{person}{F~Richard Yu}, {and} \bibinfo{person}{Victor~C Leung}.} \bibinfo{year}{2024}\natexlab{}.
\newblock \showarticletitle{Large language models (LLMs) inference offloading and resource allocation in cloud-edge computing: An active inference approach}.
\newblock \bibinfo{journal}{\emph{IEEE Transactions on Mobile Computing}} (\bibinfo{year}{2024}).
\newblock


\bibitem[Jacobs et~al\mbox{.}(1997)]%
        {jacobs1997managing}
\bibfield{author}{\bibinfo{person}{Marco~C Jacobs}, \bibinfo{person}{Mark~A Livingston}, {and} \bibinfo{person}{Andrei State}.} \bibinfo{year}{1997}\natexlab{}.
\newblock \showarticletitle{Managing latency in complex augmented reality systems}. In \bibinfo{booktitle}{\emph{Proceedings of the 1997 symposium on Interactive 3D graphics}}. \bibinfo{pages}{49--ff}.
\newblock


\bibitem[Jain(2022)]%
        {jain2022hugging}
\bibfield{author}{\bibinfo{person}{Shashank~Mohan Jain}.} \bibinfo{year}{2022}\natexlab{}.
\newblock \showarticletitle{Hugging face}.
\newblock In \bibinfo{booktitle}{\emph{Introduction to transformers for NLP: With the hugging face library and models to solve problems}}. \bibinfo{publisher}{Springer}, \bibinfo{pages}{51--67}.
\newblock


\bibitem[Kaiming~He and Sun(2015)]%
        {kaiming2015resnet}
\bibfield{author}{\bibinfo{person}{Shaoqing~Ren Kaiming~He, Xiangyu~Zhang} {and} \bibinfo{person}{Jian Sun}.} \bibinfo{year}{2015}\natexlab{}.
\newblock \showarticletitle{Deep Residual Learning for Image Recognition}.
\newblock \bibinfo{journal}{\emph{arXiv preprint}} (\bibinfo{year}{2015}).
\newblock


\bibitem[Krizhevsky et~al\mbox{.}(2017)]%
        {krizhevsky2017imagenet}
\bibfield{author}{\bibinfo{person}{Alex Krizhevsky}, \bibinfo{person}{Ilya Sutskever}, {and} \bibinfo{person}{Geoffrey~E Hinton}.} \bibinfo{year}{2017}\natexlab{}.
\newblock \showarticletitle{ImageNet classification with deep convolutional neural networks}.
\newblock \bibinfo{journal}{\emph{Commun. ACM}} \bibinfo{volume}{60}, \bibinfo{number}{6} (\bibinfo{year}{2017}), \bibinfo{pages}{84--90}.
\newblock


\bibitem[Liu et~al\mbox{.}(2024)]%
        {liu2024llm}
\bibfield{author}{\bibinfo{person}{Boyi Liu}, \bibinfo{person}{Jingwen Tong}, {and} \bibinfo{person}{Jun Zhang}.} \bibinfo{year}{2024}\natexlab{}.
\newblock \showarticletitle{Llm-slice: Dedicated wireless network slicing for large language models}. In \bibinfo{booktitle}{\emph{Proceedings of the 22nd ACM Conference on Embedded Networked Sensor Systems}}. \bibinfo{pages}{853--854}.
\newblock


\bibitem[Liu et~al\mbox{.}(2023)]%
        {liu2023llava}
\bibfield{author}{\bibinfo{person}{Haotian Liu}, \bibinfo{person}{Chunyuan Li}, \bibinfo{person}{Qingyang Wu}, {and} \bibinfo{person}{Yong~Jae Lee}.} \bibinfo{year}{2023}\natexlab{}.
\newblock \bibinfo{title}{Visual Instruction Tuning}.
\newblock


\bibitem[Mitra et~al\mbox{.}(2024)]%
        {mitra2024distributed}
\bibfield{author}{\bibinfo{person}{Purbesh Mitra}, \bibinfo{person}{Priyanka Kaswan}, {and} \bibinfo{person}{Sennur Ulukus}.} \bibinfo{year}{2024}\natexlab{}.
\newblock \showarticletitle{Distributed Mixture-of-Agents for Edge Inference with Large Language Models}.
\newblock \bibinfo{journal}{\emph{arXiv preprint arXiv:2412.21200}} (\bibinfo{year}{2024}).
\newblock


\bibitem[Nabiyouni et~al\mbox{.}(2017)]%
        {nabiyouni2017relative}
\bibfield{author}{\bibinfo{person}{Mahdi Nabiyouni}, \bibinfo{person}{Siroberto Scerbo}, \bibinfo{person}{Doug~A Bowman}, {and} \bibinfo{person}{Tobias H{\"o}llerer}.} \bibinfo{year}{2017}\natexlab{}.
\newblock \showarticletitle{Relative effects of real-world and virtual-world latency on an augmented reality training task: an ar simulation experiment}.
\newblock \bibinfo{journal}{\emph{Frontiers in ICT}}  \bibinfo{volume}{3} (\bibinfo{year}{2017}), \bibinfo{pages}{34}.
\newblock


\bibitem[Nelson and Toloui(2023)]%
        {nelson2023deploy}
\bibfield{author}{\bibinfo{person}{Sean Huver~Nigel Nelson} {and} \bibinfo{person}{Mostafa Toloui}.} \bibinfo{year}{2023}\natexlab{}.
\newblock \bibinfo{title}{Deploy Large Language Models at the Edge with NVIDIA IGX Orin Developer Kit}.
\newblock


\bibitem[Nikaein et~al\mbox{.}(2014)]%
        {nikaein2014openairinterface}
\bibfield{author}{\bibinfo{person}{Navid Nikaein}, \bibinfo{person}{Mahesh~K. Marina}, \bibinfo{person}{Saravana Manickam}, \bibinfo{person}{Alex Dawson}, \bibinfo{person}{Raymond Knopp}, {and} \bibinfo{person}{Christian Bonnet}.} \bibinfo{year}{2014}\natexlab{}.
\newblock \showarticletitle{OpenAirInterface: A Flexible Platform for 5G Research}.
\newblock \bibinfo{journal}{\emph{ACM SIGCOMM Computer Communication Review}} \bibinfo{volume}{44}, \bibinfo{number}{5} (\bibinfo{date}{October} \bibinfo{year}{2014}), \bibinfo{pages}{33--38}.
\newblock


\bibitem[NVIDIA(2025)]%
        {tensorrt_llm}
\bibfield{author}{\bibinfo{person}{NVIDIA}.} \bibinfo{year}{2025}\natexlab{}.
\newblock \bibinfo{title}{TensorRT-LLM: High-Performance Inference for Large Language Models}.
\newblock \bibinfo{howpublished}{\url{https://github.com/NVIDIA/TensorRT-LLM}}.
\newblock
\newblock
\shownote{Accessed: 2025-02-28}.


\bibitem[Ollama(2025)]%
        {ollama}
\bibfield{author}{\bibinfo{person}{Ollama}.} \bibinfo{year}{2025}\natexlab{}.
\newblock \bibinfo{title}{Ollama: An Open Source Project on GitHub}.
\newblock \bibinfo{howpublished}{\url{https://github.com/ollama/ollama}}.
\newblock
\newblock
\shownote{Accessed: 2025-02-28}.


\bibitem[Om et~al\mbox{.}(2022)]%
        {om2022h}
\bibfield{author}{\bibinfo{person}{Khandu Om}, \bibinfo{person}{Tanya McGill}, \bibinfo{person}{Michael Dixon}, \bibinfo{person}{Kok~Wai Wong}, {and} \bibinfo{person}{Polychronis Koutsakis}.} \bibinfo{year}{2022}\natexlab{}.
\newblock \showarticletitle{H. 264 and H. 265 video traffic modeling using neural networks}.
\newblock \bibinfo{journal}{\emph{Computer Communications}}  \bibinfo{volume}{184} (\bibinfo{year}{2022}), \bibinfo{pages}{149--159}.
\newblock


\bibitem[Open5GS(2025)]%
        {open5gs}
\bibfield{author}{\bibinfo{person}{Open5GS}.} \bibinfo{year}{2025}\natexlab{}.
\newblock \bibinfo{title}{Open5GS: Open Source 5G Core Network}.
\newblock \bibinfo{howpublished}{\url{https://open5gs.org/}}.
\newblock
\newblock
\shownote{Accessed: 2025-02-28}.


\bibitem[Rasley et~al\mbox{.}(2020)]%
        {rasley2020deepspeed}
\bibfield{author}{\bibinfo{person}{Jeff Rasley}, \bibinfo{person}{Samyam Rajbhandari}, \bibinfo{person}{Olatunji Ruwase}, {and} \bibinfo{person}{Yuxiong He}.} \bibinfo{year}{2020}\natexlab{}.
\newblock \showarticletitle{Deepspeed: System optimizations enable training deep learning models with over 100 billion parameters}. In \bibinfo{booktitle}{\emph{Proceedings of the 26th ACM SIGKDD international conference on knowledge discovery \& data mining}}. \bibinfo{pages}{3505--3506}.
\newblock


\bibitem[Redmon et~al\mbox{.}(2016)]%
        {redmon2016you}
\bibfield{author}{\bibinfo{person}{Joseph Redmon}, \bibinfo{person}{Santosh Divvala}, \bibinfo{person}{Ross Girshick}, {and} \bibinfo{person}{Ali Farhadi}.} \bibinfo{year}{2016}\natexlab{}.
\newblock \showarticletitle{You only look once: Unified, real-time object detection}. In \bibinfo{booktitle}{\emph{Proceedings of the IEEE conference on computer vision and pattern recognition}}. \bibinfo{pages}{779--788}.
\newblock


\bibitem[Saboorian and Xiang(2017)]%
        {saboorian2017network}
\bibfield{author}{\bibinfo{person}{Tony Saboorian} {and} \bibinfo{person}{Amanda Xiang}.} \bibinfo{year}{2017}\natexlab{}.
\newblock \showarticletitle{Network Slicing and 3GPP Service and Systems Aspects (SA) Standard}.
\newblock \bibinfo{journal}{\emph{IEEE Software Defined Networks}} (\bibinfo{date}{December} \bibinfo{year}{2017}).
\newblock


\bibitem[Schmidt et~al\mbox{.}(2021)]%
        {schmidt2021flexric}
\bibfield{author}{\bibinfo{person}{Robert Schmidt}, \bibinfo{person}{Mikel Irazabal}, {and} \bibinfo{person}{Navid Nikaein}.} \bibinfo{year}{2021}\natexlab{}.
\newblock \showarticletitle{FlexRIC: An SDK for next-generation SD-RANs}. In \bibinfo{booktitle}{\emph{Proceedings of the 17th International Conference on emerging Networking EXperiments and Technologies}}. \bibinfo{pages}{411--425}.
\newblock


\bibitem[Stojkovic et~al\mbox{.}(2025)]%
        {stojkovic2025dynamollm}
\bibfield{author}{\bibinfo{person}{Jovan Stojkovic}, \bibinfo{person}{Chaojie Zhang}, \bibinfo{person}{{\'I}{\~n}igo Goiri}, \bibinfo{person}{Josep Torrellas}, {and} \bibinfo{person}{Esha Choukse}.} \bibinfo{year}{2025}\natexlab{}.
\newblock \showarticletitle{Dynamollm: Designing llm inference clusters for performance and energy efficiency}. In \bibinfo{booktitle}{\emph{2025 IEEE International Symposium on High Performance Computer Architecture (HPCA)}}. IEEE, \bibinfo{pages}{1348--1362}.
\newblock


\bibitem[Tang et~al\mbox{.}(2023)]%
        {tang2023tinychat}
\bibfield{author}{\bibinfo{person}{Haotian Tang}, \bibinfo{person}{Shang Yang}, \bibinfo{person}{Ji Lin}, \bibinfo{person}{Jiaming Tang}, \bibinfo{person}{Wei-Ming Chen}, \bibinfo{person}{Wei-Chen Wang}, {and} \bibinfo{person}{Song Han}.} \bibinfo{year}{2023}\natexlab{}.
\newblock \showarticletitle{TinyChat: Large Language Model on the Edge}.
\newblock \bibinfo{journal}{\emph{MIT HAN Lab Blog}} (\bibinfo{year}{2023}).
\newblock
\urldef\tempurl%
\url{https://hanlab.mit.edu/blog/tinychat}
\showURL{%
\tempurl}


\bibitem[Wu et~al\mbox{.}(2020)]%
        {wu2020deep}
\bibfield{author}{\bibinfo{person}{Huaming Wu}, \bibinfo{person}{Xiangyi Li}, {and} \bibinfo{person}{Yingjun Deng}.} \bibinfo{year}{2020}\natexlab{}.
\newblock \showarticletitle{Deep learning-driven wireless communication for edge-cloud computing: opportunities and challenges}.
\newblock \bibinfo{journal}{\emph{Journal of Cloud Computing}} \bibinfo{volume}{9}, \bibinfo{number}{1} (\bibinfo{year}{2020}), \bibinfo{pages}{21}.
\newblock


\bibitem[Xu et~al\mbox{.}(2025)]%
        {xu2025serving}
\bibfield{author}{\bibinfo{person}{Minrui Xu}, \bibinfo{person}{Dusit Niyato}, {and} \bibinfo{person}{Christopher~G Brinton}.} \bibinfo{year}{2025}\natexlab{}.
\newblock \showarticletitle{Serving Long-Context LLMs at the Mobile Edge: Test-Time Reinforcement Learning-based Model Caching and Inference Offloading}.
\newblock \bibinfo{journal}{\emph{arXiv preprint arXiv:2501.14205}} (\bibinfo{year}{2025}).
\newblock


\bibitem[Xue et~al\mbox{.}(2024)]%
        {xue2024wdmoe}
\bibfield{author}{\bibinfo{person}{Nan Xue}, \bibinfo{person}{Yaping Sun}, \bibinfo{person}{Zhiyong Chen}, \bibinfo{person}{Meixia Tao}, \bibinfo{person}{Xiaodong Xu}, \bibinfo{person}{Liang Qian}, \bibinfo{person}{Shuguang Cui}, \bibinfo{person}{Wenjun Zhang}, {and} \bibinfo{person}{Ping Zhang}.} \bibinfo{year}{2024}\natexlab{}.
\newblock \showarticletitle{WDMoE: Wireless Distributed Mixture of Experts for Large Language Models}.
\newblock \bibinfo{journal}{\emph{arXiv preprint arXiv:2411.06681}} (\bibinfo{year}{2024}).
\newblock


\bibitem[Zhang et~al\mbox{.}(2025)]%
        {zhang2025beyond}
\bibfield{author}{\bibinfo{person}{Xiaoyu Zhang} {et~al\mbox{.}}} \bibinfo{year}{2025}\natexlab{}.
\newblock \showarticletitle{Beyond the Cloud: Edge Inference for Generative Large Language Models in Wireless Networks}.
\newblock \bibinfo{journal}{\emph{IEEE Transactions on Wireless Communications}} (\bibinfo{year}{2025}).
\newblock


\bibitem[Zhao et~al\mbox{.}(2019)]%
        {zhao2019object}
\bibfield{author}{\bibinfo{person}{Zhong-Qiu Zhao}, \bibinfo{person}{Peng Zheng}, \bibinfo{person}{Shou-tao Xu}, {and} \bibinfo{person}{Xindong Wu}.} \bibinfo{year}{2019}\natexlab{}.
\newblock \showarticletitle{Object detection with deep learning: A review}.
\newblock \bibinfo{journal}{\emph{IEEE transactions on neural networks and learning systems}} \bibinfo{volume}{30}, \bibinfo{number}{11} (\bibinfo{year}{2019}), \bibinfo{pages}{3212--3232}.
\newblock


\bibitem[Zheng et~al\mbox{.}(2024)]%
        {zheng2024review}
\bibfield{author}{\bibinfo{person}{Yue Zheng}, \bibinfo{person}{Yuhao Chen}, \bibinfo{person}{Bin Qian}, \bibinfo{person}{Xiufang Shi}, \bibinfo{person}{Yuanchao Shu}, {and} \bibinfo{person}{Jiming Chen}.} \bibinfo{year}{2024}\natexlab{}.
\newblock \showarticletitle{A Review on edge large language models: Design, Execution, and Applications}.
\newblock \bibinfo{journal}{\emph{Comput. Surveys}} (\bibinfo{year}{2024}).
\newblock


\end{thebibliography}
\appendix
\begin{onecolumn}
\section*{\LARGE Appendix}
\section{Appendix Overview}
This appendix provides supplementary information, analysis, and technical details to support the main body of the paper. The contents are organized as follows:

Appendix~\ref{appendixBackground} presents a general comparison of WiLLM against existing open-source wireless communication systems and LLM deployment frameworks. This comparative analysis establishes WiLLM's distinctive contributions within the telecommunications-AI convergence landscape.

Appendix~\ref{appendixBackground2} extends the comparative analysis through a granular examination of specific technical capabilities across contemporary wireless systems. This detailed evaluation substantiates WiLLM's position as the first comprehensive system specifically designed for wireless LLM service delivery.

Appendix~\ref{appendixSystem} illustrates the physical implementation of the WiLLM testbed, depicting the comprehensive hardware architecture that enables end-to-end LLM service delivery over wireless networks. It also showcases the WiLLM monitoring and performance analysis GUI, which provides real-time visualization and analytical capabilities for system optimization.

Appendix~\ref{appendixSystemAPI} details WiLLM's hierarchical API architecture and presents the Tree-Branch-Fruit Slicing algorithm for UE resource allocation. These technical specifications elucidate the system's cross-layer integration mechanisms and resource distribution methodologies.

Appendix~\ref{appendixDataset} contains supplementary data analysis and visualizations from the WiLLM dataset. These figures provide deeper insights into the stability of experimental conditions, comparative performance between slice-enabled and normal traffic, and the complex interactions between various system parameters. More importantly, we conducted a causal analysis of the results shown in Figure.~\ref{fig:Uplink_UE_Dynamic} to Figure ~\ref{fig:Slice_bytes} of the dataset.

Appendix~\ref{benchmark} introduces two novel benchmarking metrics - the LLM-Aware Resource Efficiency Index (LAREI) and the LLM Slice Efficiency Quotient (LSEQ). These metrics are derived from rigorous analysis of the WiLLM dataset and provide a comprehensive framework for evaluating the performance of LLM wireless communication systems.

Appendix~\ref{appendixMatrics} presents a comprehensive tabulation of all metrics collected in the WiLLM dataset across the UE, RAN, and Edge Server layers. This exhaustive catalog enables researchers to understand the full scope and granularity of the empirical data foundation provided by WiLLM.

The appendix aims to provide a rigorous technical foundation for the innovative architectural designs and empirical findings presented in the main paper. By offering detailed comparative analysis, comprehensive system specifications, in-depth dataset insights, and novel performance evaluation frameworks, this supplementary material enables researchers to fully engage with the technical contributions of WiLLM and build upon its open-source foundation to advance the state of the art in wireless LLM service delivery.

\clearpage
\section{Appendix for Section~\ref{sec:related} (Related Works): General Comparation of Systems}
\label{appendixBackground}
\definecolor{lightblue}{RGB}{135,206,250}
\definecolor{lightred}{RGB}{255,182,193}
\begin{table*}[htbp]
\centering
\renewcommand{\tabularxcolumn}[1]{m{#1}}
\caption{Comparison of Open Source Wireless Communication Systems and LLM deployment Systems}
\label{ComparationSystems}
\begin{tabularx}{1\textwidth}{%
>{\bfseries\centering\arraybackslash}X  
>{\centering\arraybackslash}X 
>{\centering\arraybackslash}X 
>{\centering\arraybackslash}X 
>{\centering\arraybackslash}X 
>{\centering\arraybackslash}X 
>{\centering\arraybackslash}X 
>{\centering\arraybackslash}X
>{\centering\arraybackslash}X}
\toprule
\textbf{System Name} & \textbf{Network Architecture} & \textbf{Slice Support} & \textbf{LLM Integration} & \textbf{Resource Scheduling} & \textbf{UE Compatibility} & \textbf{API Framework} & \textbf{Deployment Flexibility} & \textbf{Dataset \& Benchmark} \\
\midrule
WiLLM & \textcolor{blue}{UE-gNB-CN with Edge Server} & \textcolor{blue}{``Tree-Branch-Fruit'' Slice} & \textcolor{blue}{Designed specifically for LLM services} & \textcolor{blue}{Dual-mode resource schedule} & \textcolor{blue}{All UE} & \textcolor{blue}{Cross-layer API framework} & \textcolor{blue}{Core and edge part flexibility} & \textcolor{blue}{\cmark} \\
\midrule
OAI \cite{nikaein2014openairinterface} & \textcolor{blue}{UE-gNB-CN} & \textcolor{blue}{Base slice} & \textcolor{blue}{\xmark} & \textcolor{blue}{Basic resource schedule} & \textcolor{blue}{Slice-enabled UE} & \textcolor{blue}{gNB Only} & \textcolor{blue}{Traditional service-oriented} & \textcolor{blue}{\xmark} \\
\midrule
srsRAN \cite{flakowski2023implementation} & \textcolor{blue}{UE-gNB-CN} & \textcolor{blue}{Base slice} & \textcolor{blue}{\xmark} & \textcolor{blue}{Flexible resource schedule} & \textcolor{blue}{Slice-enabled UE} & \textcolor{blue}{gNB Only} & \textcolor{blue}{Traditional service-oriented} & \textcolor{blue}{\xmark} \\
\midrule
Open5GS \cite{open5gs} & \textcolor{blue}{UE-gNB-CN} & \textcolor{blue}{Base slice} & \textcolor{blue}{\xmark} & \textcolor{blue}{Basic resource schedule} & \textcolor{blue}{Slice-enabled UE} & \textcolor{blue}{gNB Only} & \textcolor{blue}{Traditional service-oriented} & \textcolor{blue}{\xmark} \\
\midrule
Hugging Face's TGI \cite{jain2022hugging} &  LLM only & \xmark & Fine-tuning and optimization & \xmark & \xmark & Computation Only & Server & \xmark \\
\midrule
Ollama \cite{ollama} &  LLM only & \xmark & LLM deployment & \xmark & \xmark & Computation Only & Server & \xmark \\
\midrule
TensorRT-LLM \cite{tensorrt_llm} &  LLM only & \xmark & Lightweight reasoning & \xmark & \xmark & Computation Only & Server & \xmark \\
\midrule
DeepSpeed \cite{rasley2020deepspeed} &  LLM only & \xmark & Distributed training and reasoning & \xmark & \xmark & Computation Only & Server & \xmark \\
\bottomrule
\end{tabularx}
{\footnotesize
~\\
\textbf{Legend:} \cmark~Support~~\xmark~Not Considered~~{\color{blue}{Blue text:}}~key comparisons\\
\vspace{0.1cm}
\parbox{\textwidth}{\justifying
\textbf{Note:} This comparison contrasts systems with different foundations: OAI, srsRAN, and Open5GS are ground-up implementations, while WiLLM strategically extends OAI's architecture. While this may appear uneven, the comparison serves to highlight functional gaps in existing systems and how WiLLM addresses these through targeted extensions. This approach enables rapid innovation while leveraging established frameworks, demonstrating how focused architectural evolution can efficiently address emerging application needs.
}}
\end{table*}
Table~\ref{ComparationSystems} presents a comprehensive comparative analysis of WiLLM against existing open-source wireless communication systems and LLM deployment frameworks. This systematic evaluation establishes WiLLM's distinctive contributions within the telecommunications-AI convergence landscape. While traditional systems (OAI, srsRAN, Open5GS) provide foundational wireless infrastructure components, they exhibit significant limitations in LLM integration capabilities. Concurrently, established LLM frameworks (Hugging Face's TGI, Ollama, TensorRT-LLM, DeepSpeed) offer sophisticated model deployment but lack wireless communication integration. WiLLM uniquely bridges this interdisciplinary gap through its "Tree-Branch-Fruit" slicing architecture, cross-layer API framework, and universal UE compatibility mechanisms. This comparative evaluation demonstrates how WiLLM transcends the limitations of existing systems through its specialized design for LLM service delivery in wireless environments.
\end{onecolumn}
\clearpage

\section{Appendix for Section~\ref{sec:related} (Related Works): Detailed Technical Comparison of Systems}
\label{appendixBackground2}
\definecolor{green}{RGB}{34,139,34}
\definecolor{red}{RGB}{255,0,0}
\definecolor{yellow}{RGB}{255,191,0}
\newcolumntype{x}{>{\centering\arraybackslash}p{1.7cm}}
\newcommand{\greenmark}{\textcolor{green}{\ding{110}}}
\newcommand{\redmark}{\textcolor{red}{\ding{110}}}
\newcommand{\yellowmark}{\textcolor{yellow}{\ding{110}}}
\begin{table*}[htbp]
\centering
\footnotesize
\setlength{\tabcolsep}{4pt}
\caption{Technical Comparison of Wireless Systems}
\label{benchmarkComTable}
\begin{tabular}{p{4.5 cm}*{6}{x}}
\toprule
\textbf{Technical Features} & \textbf{WiLLM} & \textbf{OAI} & \textbf{srsRAN} & \textbf{Open5GS} & \textbf{FlexRIC} & \textbf{Janus} \\
\midrule
\textbf{Type} & LLM-Oriented Platform & Open-Source 5G System & Open-Source 4G/5G Suite & Open-Source 5G CN & Programmable RAN SDK & Programmable RAN Platform \\
\midrule
\textbf{Main Features} & Specialized for LLM Comm. & Research and Prototyping & eNodeB and UE Apps & LTE/NR Core Network & xAPP Development & O-RAN Service Model \\
\midrule
LLM-Specific Slicing Architecture & \greenmark & \redmark & \redmark & \redmark & \redmark & \redmark \\
Dynamic Slice Compatibility & \greenmark & \redmark & \redmark & \redmark & \yellowmark & \yellowmark \\
Universal UE Compatibility for Slice & \greenmark & \redmark & \redmark & \redmark & \yellowmark & \yellowmark \\
Multi-UE-Multi-Slice Coordination & \greenmark & \redmark & \redmark & \redmark & \yellowmark & \yellowmark \\
Dual-Mode Resource Scheduling & \greenmark & \redmark & \redmark & \redmark & \yellowmark & \yellowmark \\
Cross-Layer API Framework & \greenmark & \redmark & \redmark & \redmark & \yellowmark & \yellowmark \\
Flexible LLM Deployment & \greenmark & \redmark & \redmark & \redmark & \redmark & \redmark \\
LLM Communication Dataset & \greenmark & \redmark & \redmark & \redmark & \redmark & \redmark \\
LLM Communication Benchmark & \greenmark & \redmark & \redmark & \redmark & \redmark & \redmark \\
Hierarchical Slice Policy Enforcement & \greenmark & \redmark & \redmark & \redmark & \yellowmark & \yellowmark \\
Application-Layer Slice Access & \greenmark & \redmark & \redmark & \redmark & \yellowmark & \yellowmark \\
Synchronized Multi-Interface Metrics & \greenmark & \redmark & \redmark & \redmark & \yellowmark & \yellowmark \\
Offline and Online Slice Optimization & \greenmark & \redmark & \redmark & \redmark & \yellowmark & \yellowmark \\
Programmability Interface & \greenmark & \redmark & \redmark & \redmark & \greenmark & \greenmark \\
xApp/rApp Support & \yellowmark & \redmark & \redmark & \redmark & \greenmark & \greenmark \\
Disaggregated RAN Architecture & \yellowmark & \yellowmark & \yellowmark & \redmark & \greenmark & \greenmark \\
RAN Intelligence Controller Integration & \yellowmark & \redmark & \redmark & \redmark & \greenmark & \greenmark \\
Service Model Flexibility & \yellowmark & \redmark & \redmark & \redmark & \yellowmark & \greenmark \\
Multi-vendor RAN Integration & \redmark & \redmark & \redmark & \redmark & \greenmark & \greenmark \\
Low-latency Optimization Tools & \greenmark & \yellowmark & \yellowmark & \redmark & \greenmark & \yellowmark \\
O-RAN Compliance & \yellowmark & \redmark & \redmark & \redmark & \greenmark & \greenmark \\
Protocol Strictness & \redmark & \greenmark & \greenmark & \greenmark & \yellowmark & \yellowmark \\
Original Protocol Stack Implementation & \redmark & \greenmark & \greenmark & \greenmark & \yellowmark & \yellowmark \\
Original Hardware Support & \greenmark & \greenmark & \greenmark & \greenmark & \yellowmark & \yellowmark \\
Original System Architecture & \yellowmark & \greenmark & \greenmark & \greenmark & \yellowmark & \yellowmark \\
Open-Source Community Activity & \yellowmark & \greenmark & \greenmark & \greenmark & \yellowmark & \yellowmark \\
Standard Compliance Testing Tools & \redmark & \greenmark & \greenmark & \yellowmark & \yellowmark & \redmark \\
Cost Efficiency for Deployment & \yellowmark & \greenmark & \greenmark & \greenmark & \yellowmark & \yellowmark \\
\bottomrule
\end{tabular}
\vspace{0.05cm}
\\
{\footnotesize
\textbf{Legend:} \greenmark~Full support/Good~~\redmark~Not supported/Bad~~\yellowmark~Partial support/Mediocre
\\
\textbf{Note:} WiLLM is an extend of OAI}
\label{tab:system_comparison}
\end{table*}
Table~\ref{benchmarkComTable} extends our comparative analysis through a granular examination of specific technical capabilities across contemporary wireless systems. This evaluation reveals significant limitations in existing platforms regarding LLM service integration. While established systems like OAI, srsRAN, and Open5GS offer partial implementation of core wireless networking features, they demonstrate substantial deficiencies in specialized LLM communication capabilities. WiLLM distinguishes itself through full implementation of LLM-specific features, including dynamic slice compatibility, multi-UE-multi-slice coordination, and hierarchical slice policy enforcement. The matrix also highlights WiLLM's innovative application-layer slice access mechanism, which enables universal device compatibility without requiring protocol-level modifications. This detailed technical comparison substantiates WiLLM's position as the first comprehensive system specifically designed for wireless LLM service delivery.
\clearpage

\section{Appendix for Section~\ref{sec:implementation} (System): The Hardware Platform}
\label{appendixSystem}
\begin{figure*}[htbp]
    \centering
    \includegraphics[width=0.98\textwidth]{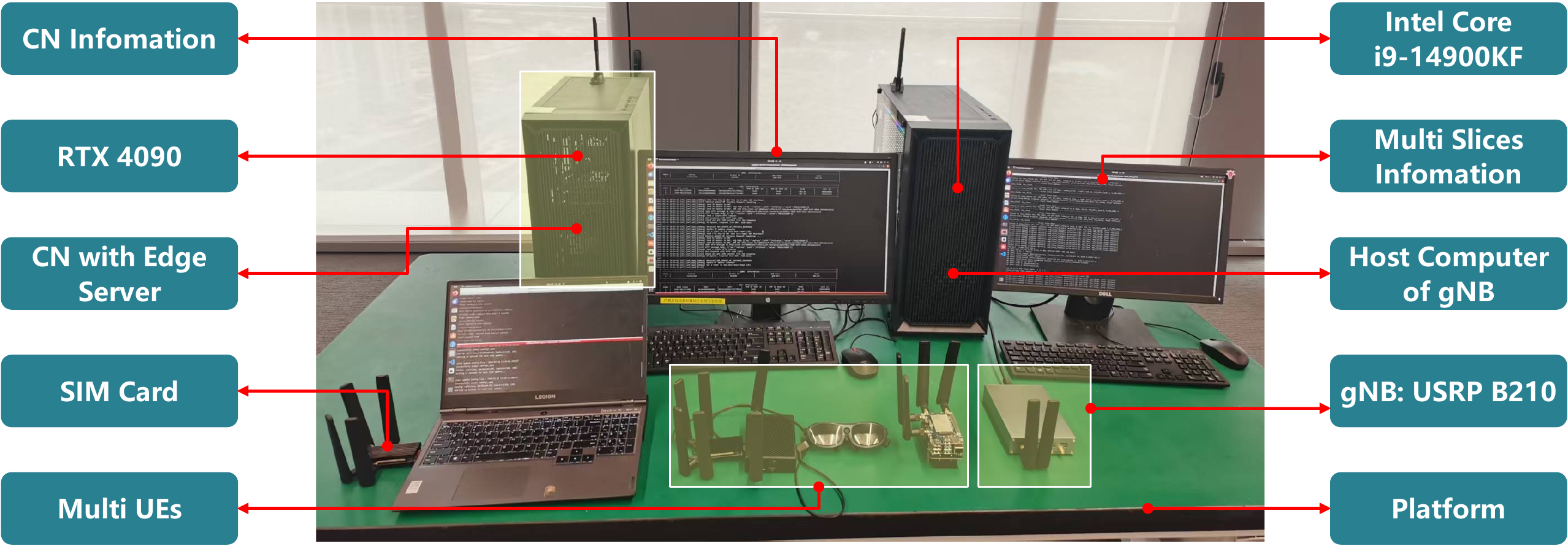}
    \caption{Hardware deployment of the WiLLM testbed, showcasing the integrated system with core network servers, edge computing nodes, Radio Access Network equipment, and multiple user terminal devices.}
    \label{fig:platform}
\end{figure*}
Figure~\ref{fig:platform} depicts the physical implementation of our WiLLM testbed, illustrating the comprehensive hardware architecture that enables end-to-end LLM service delivery over wireless networks. The deployment integrates multiple hierarchical components: an Intel Core i9-14900KF processor serving as the gNB host, a USRP B210 software-defined radio functioning as the RF frontend, and an NVIDIA RTX 4090 GPU enabling efficient LLM inference processing. The system supports simultaneous multi-slice and multi-UE operations through specialized hardware components and SIM card integration. This hardware configuration establishes a complete signal chain from core network to edge computing infrastructure, facilitating comprehensive experimental validation of the WiLLM architecture across diverse deployment scenarios and computational requirements.
\begin{figure}[htbp]
    \centering
    \includegraphics[width=0.7\textwidth]{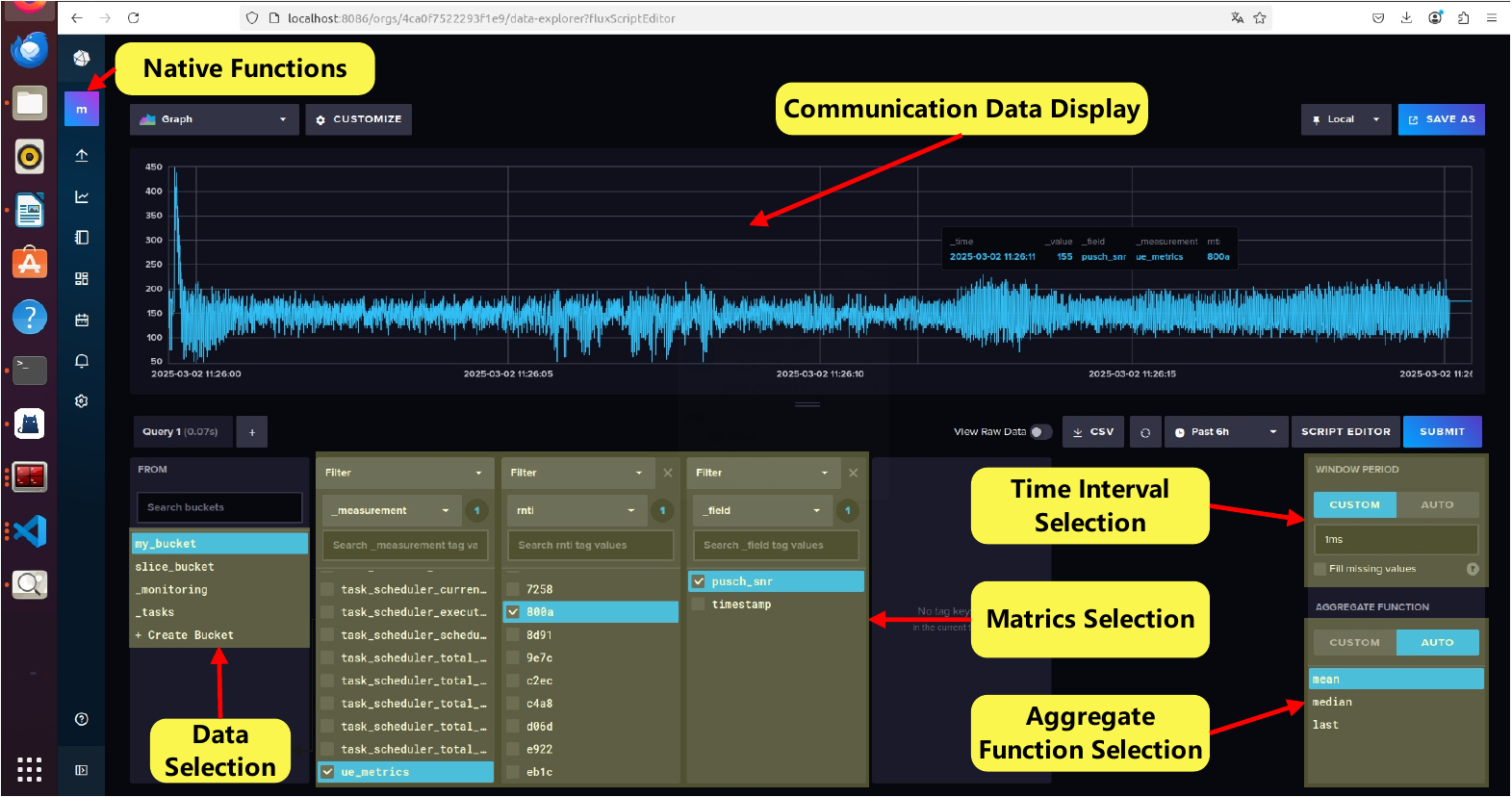}
    \caption{WiLLM monitoring and performance analysis GUI, supporting multi-dimensional metric visualization, time-series analysis, and aggregate function computation.}
    \label{fig:GUI}
\end{figure}
\\
Figure~\ref{fig:GUI} illustrates the WiLLM monitoring and performance analysis GUI, which provides real-time visualization and analytical capabilities for system optimization. The interface implements a multi-dimensional metric visualization framework supporting time-series analysis and aggregate function computation. Key functional components include: (1) communication data display for real-time performance monitoring, (2) metrics selection interface for parameter filtering and comparison, (3) time interval selection for temporal analysis, (4) aggregate function selector for statistical computation, and (5) data selection mechanisms for customizing visualization parameters. This integrated monitoring framework enables comprehensive performance evaluation across multiple system layers, facilitating both runtime optimization and longitudinal performance analysis.
\clearpage

\begin{twocolumn}
\section{Appendix for Section~\ref{sec:implementation} (System): The API Architecture and Slice Algorithm of WiLLM}
\label{appendixSystemAPI}
\begin{figure}[htbp]
    \centering
    \includegraphics[width=0.48\textwidth]{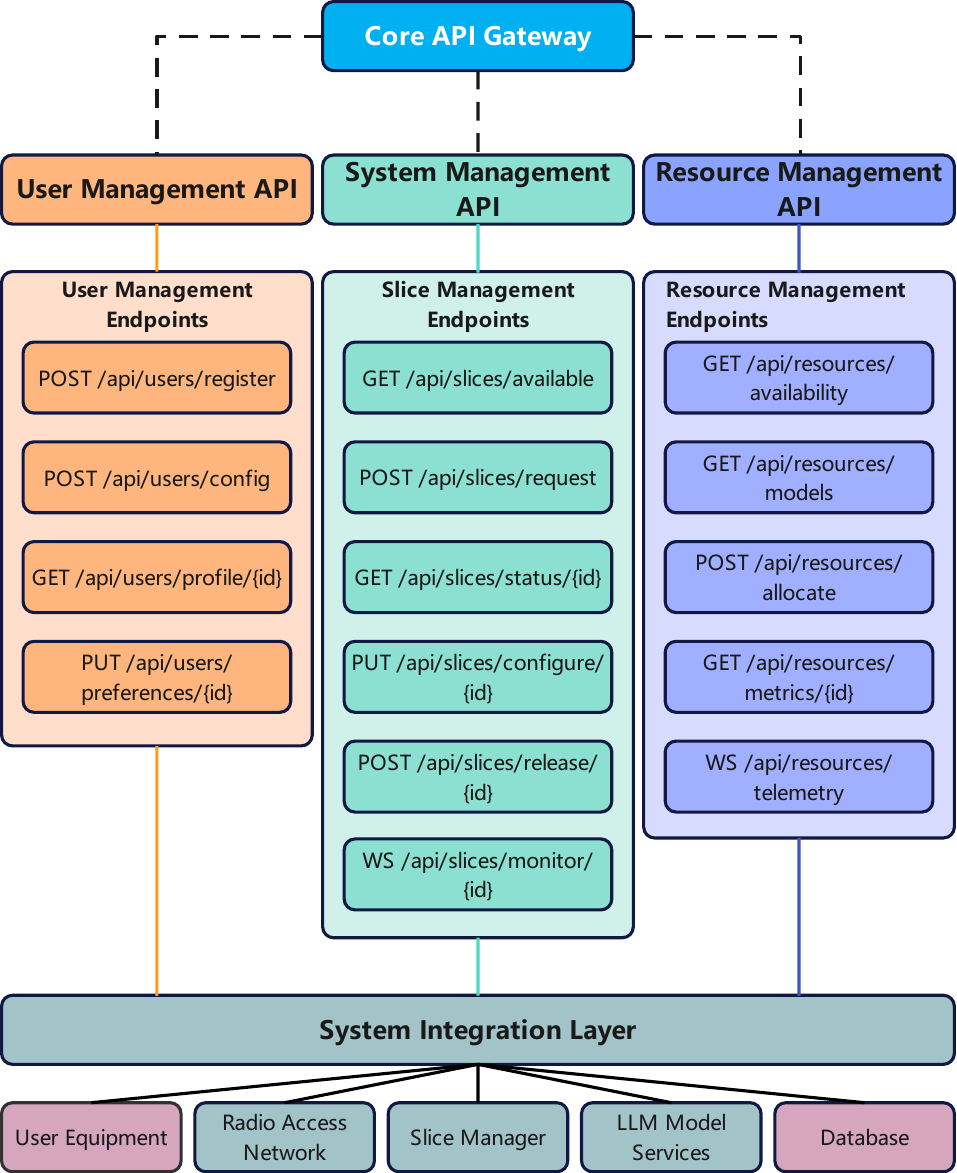}
    \caption{API Architecture of WiLLM.}
    \label{fig:api}
\end{figure}
Figure~\ref{fig:api} illustrates WiLLM's hierarchical API architecture with a three-tier structure: (1) User Management API handling registration, configuration and preferences; (2) System Management API orchestrating slice operations through availability checks, request processing, and status monitoring; and (3) Resource Management API managing resource discovery, allocation and telemetry. This architecture implements cross-layer integration while maintaining separation of concerns, using RESTful APIs for synchronous operations and WebSocket connections for asynchronous monitoring.

Algorithm~\ref{alg:ue_slice_math} presents the Tree-Branch-Fruit Slicing algorithm for UE resource allocation. It processes UE identity, PRBs, slice policies, channel quality indicators, and throughput metrics through four phases: (1) branch slice determination using NSSAI configuration (lines 1-4); (2) transport block size computation with proportional fair metrics (lines 5-7); (3) hierarchical policy enforcement across branches and fruits (lines 8-13); and (4) final resource allocation with MCS selection (lines 14-16). This implementation ensures efficient resource distribution while maintaining slice isolation and service differentiation across multiple UEs.
\begin{algorithm}[t]
\caption{Tree–Branch–Fruit Slicing for UEs}
\label{alg:ue_slice_math}
\begin{algorithmic}[1]
\REQUIRE UE $u$, total PRBs $N_{\text{PRB}}$, slice policies $\mathcal{P}$, channel quality parameters $Q_m,\, R,\, n_\text{RB},\, n_\text{sym},\, L$, and historical throughput $\Theta(u)$.
\ENSURE Final allocated PRB count $R(u)$ and selected MCS for $u$.

\STATE \textbf{/* Determine UE's branch slice based on NSSAI configuration */}
\STATE Define the slice set of $u$ as 
\[
\mathcal{S}(u) = \{\, s \mid s \text{ is extracted from } \mathrm{NSSAI}(u) \,\}
\]
\STATE Compute the branch slice index 
\[
b = \operatorname{MatchBranch}(\mathcal{S}(u), \mathcal{P})
\]
\STATE Retrieve branch policy parameters:
\[
\alpha_b^{\min} = \mathcal{P}(b).\min\_ratio \quad \& \quad \alpha_b^{\max} = \mathcal{P}(b).\max\_ratio
\]

\STATE \textbf{/* Compute Transport Block Size using channel parameters */}
\[
\text{TBS}(u) = f\bigl(Q_m,\, R,\, n_\text{RB},\, n_\text{sym},\, L\bigr)
\]
\STATE Compute the proportional fair metric:
\[
\gamma(u) = \frac{\text{TBS}(u)}{\Theta(u)}
\]
\STATE Derive the initial allocation proportion via a function $\phi(\cdot)$:
\[
r_{\text{init}} = N_{\text{PRB}} \cdot \phi\bigl(\gamma(u)\bigr)
\]
\STATE Enforce branch policy constraints (lower and upper):
\[
r_{\text{branch}} = \min\Bigl\{\, r_{\text{init}},\; N_{\text{PRB}} \cdot \alpha_b^{\max} \Bigr\}
\]
\[
r_{\text{branch}} = \max\Bigl\{\, r_{\text{branch}},\; N_{\text{PRB}} \cdot \alpha_b^{\min} \Bigr\}
\]

\IF{a fruit mapping exists for $u$}
  \STATE Let $(\pi(u),\; r_{\min},\; r_{\max})$ be the fruit parameters for $u$.
\ELSE
  \STATE Define default fruit parameters:
  \[
  \pi(u)=1,\quad r_{\min} = N_{\text{PRB}} \cdot \alpha_b^{\min},\quad r_{\max} = N_{\text{PRB}} \cdot \alpha_b^{\max}
  \]
\ENDIF
\STATE Final resource allocation:
\[
R(u) = \min\Bigl\{\, \max\Bigl\{\, \pi(u) \cdot r_{\text{branch}},\; r_{\min} \Bigr\},\; r_{\max} \Bigr\}
\]

\STATE Determine MCS from channel conditions:
\[
\text{MCS} = \operatorname{SelectMCS}(u, Q_m, R, L)
\]

\RETURN $(R(u),\, \text{MCS})$
\end{algorithmic}
\end{algorithm}
\end{twocolumn}

\begin{onecolumn}
\section{Appendix for Section~\ref{sec:evaluation} (Dataset)}
\label{appendixDataset}

\begin{figure*}[!htbp]
    \centering
    \includegraphics[width=1\textwidth]{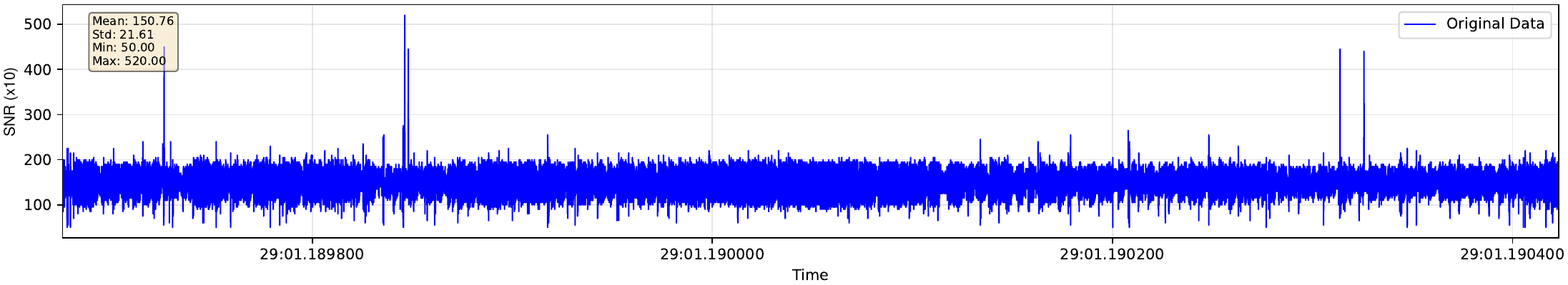}
    \caption{The variations in SNR during the data collection process.}
    \label{fig:SNR}
\end{figure*}
\begin{figure*}[!htbp]
    \centering
    \includegraphics[width=1\textwidth]{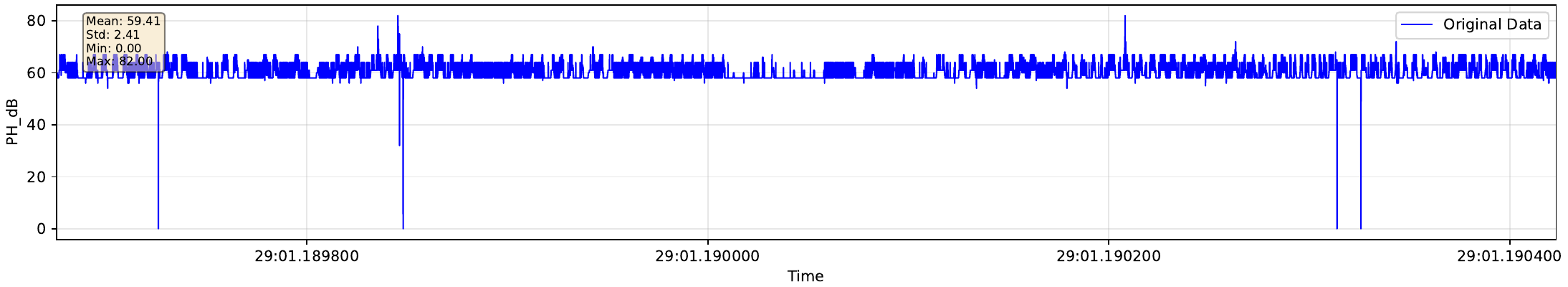}
    \caption{The variations in PH\_dB during the data collection process.}
    \label{fig:PH_dB}
\end{figure*}
\begin{figure*}[!htbp]
    \centering
    \includegraphics[width=1\textwidth]{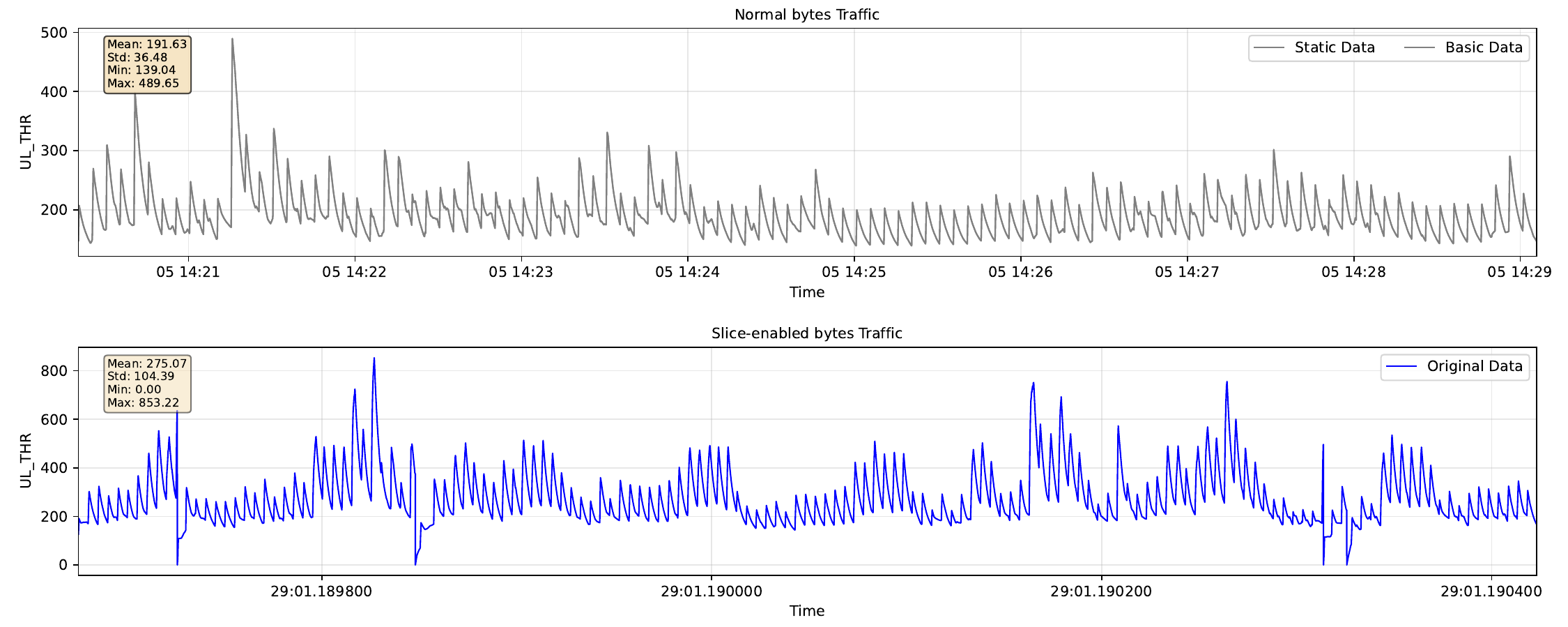}
    \caption{The variations in UL\_THR during the data collection process.}
    \label{fig:UL_THR}
\end{figure*}
Figures~\ref{fig:SNR} and ~\ref{fig:PH_dB} present temporal analyses of Signal-to-Noise Ratio (SNR) and Power Headroom (PH\_dB) variations during the data collection process, respectively. These visualizations demonstrate the stability of the experimental environment across multiple measurement dimensions. Figure~\ref{fig:SNR} reveals consistent SNR levels (mean: 150.76, standard deviation: 21.61) across the data collection period, with minimal transient fluctuations, confirming reliable signal quality for experimental validity. Similarly, Figure~\ref{fig:PH_dB} shows highly stable power headroom measurements (mean: 59.41, standard deviation: 2.41), indicating consistent UE transmission capabilities throughout the experimental process. These stability metrics validate the integrity of our dataset by confirming that the observed performance variations result from intentional experimental manipulations rather than environmental fluctuations or hardware inconsistencies.

Figure~\ref{fig:UL_THR} displays comparative analyses of uplink throughput between normal traffic (top) and slice-enabled traffic (bottom) during the data collection process. The slice-enabled implementation demonstrates significantly higher average throughput compared to normal traffic, representing a 43.5\% performance improvement. The slice-enabled approach also shows more dynamic adaptation with higher peak throughput, enabling better responsiveness to varying service demands. While normal traffic exhibits regular but constrained patterns, the slice-enabled traffic shows more variable but overall superior performance characteristics with distinct periodic bursts that effectively utilize available bandwidth. This empirical evidence confirms that WiLLM's slicing architecture successfully enhances throughput efficiency, allowing more responsive performance scaling for LLM services while maintaining an appropriate baseline throughput level, which is crucial for consistent user experience in latency-sensitive LLM applications.


\subsection{Dataset Collection Parameters}
\subsubsection{UE Configuration Parameters}
During data collection, we employed a configurable UE client with the following parameter ranges:
\begin{table}[ht]
\centering
\begin{tabular}{lp{6.5cm}p{4.5cm}}
\toprule
\textbf{Parameter} & \textbf{Description} & \textbf{Values} \\
\midrule
Image Capture Resolution & Resolution of images captured by smart glasses & 320$\times$240 to 640$\times$480 pixels \\
Display Resolution & Resolution of the glasses display panel & 800$\times$600 to 1280$\times$720 pixels \\
Request Mode & Type of request sent to LLM & ``image\_request'', ``text\_request'' \\
LLM Model & Model used for inference & ``llava'', ``llama3.2'' \\
Response Word Count & Maximum words requested in LLM response & 50, 100, 150, 200 words \\
Request Frequency & Periodic request interval & 5 seconds (default) \\
\bottomrule
\end{tabular}
\caption{UE configuration parameters used in dataset collection}
\label{tab:ue_params}
\end{table}

Our experiment controller dynamically adjusted these parameters with varying coefficients (1.0, 0.9, 0.8, 0.7, 0.6, 0.5) applied to the base resolutions to analyze performance across different capture qualities. This approach simulated diverse real-world usage scenarios with varying resource constraints and quality requirements.

\subsubsection{Network Slicing Configuration}
The network slicing configuration used in our experiments consisted of three primary slices with different resource allocation parameters. Three fruit slices were defined with max\_ratio values of \(\{30\%, 60\%, 90\%\}\), all associated with the first parent slice. During data collection, our slice controller cycled through these configurations at 30-second intervals, updating the slice mapping for UEs to create varied resource allocation patterns throughout the dataset collection process.
\end{onecolumn}

\section{Benchmarking Metrics for LLM Wireless Communication}
\label{benchmark}
\begin{figure}[htbp]
    \centering
    \begin{minipage}[t]{0.47\textwidth}
        \centering
        \includegraphics[width=\textwidth]{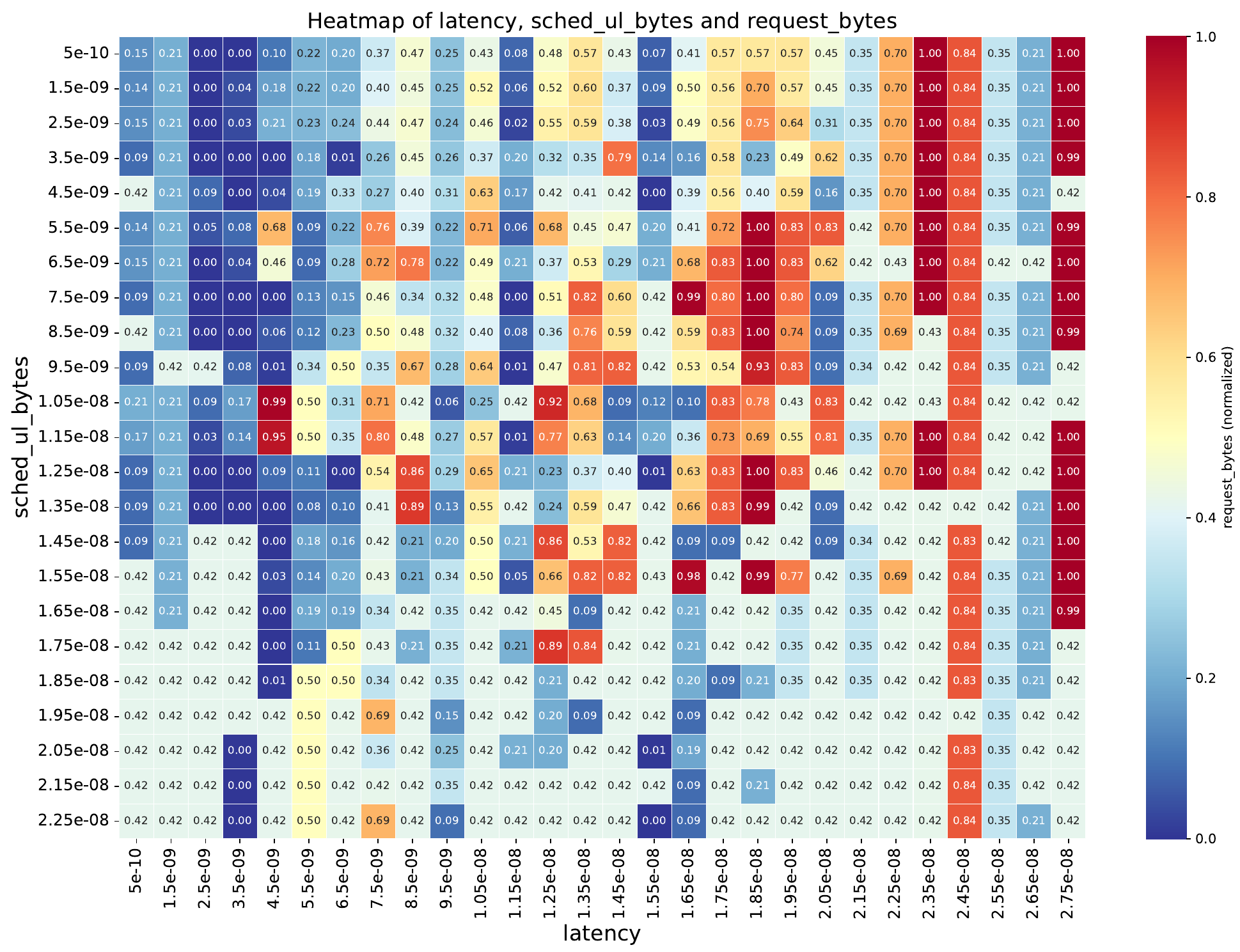}
        \caption{Latency, scheduled uplink bytes, and request bytes correlation heatmap. This figure presents a heatmap visualization of the complex interdependencies between latency, scheduled uplink bytes, and request bytes. The color-coded matrix reveals non-linear relationships between these parameters, identifying potential optimization opportunities and constraint boundaries that inform the development of comprehensive performance metrics for LLM wireless systems.}
        \label{fig:Compre_heat}
    \end{minipage}
    \hfill 
    \begin{minipage}[t]{0.5\textwidth}
        \centering
        \includegraphics[width=\textwidth]{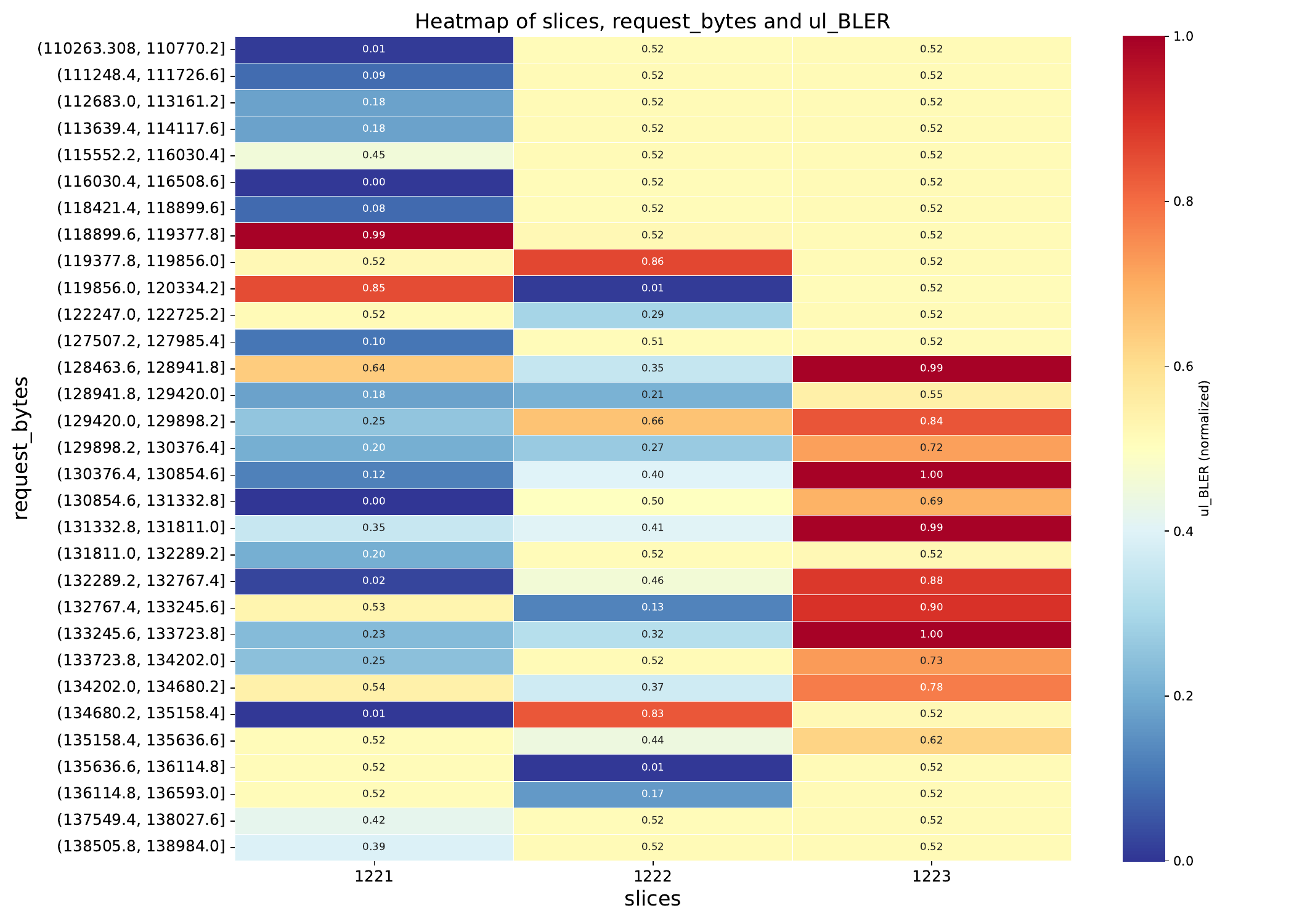}
        \caption{Correlation heatmap between slice configuration, request bytes, and uplink BLER. This figure presents a multi-dimensional analysis of how slice configuration, request payload size, and resulting error rates interact. The color-coded visualization provides empirical evidence for non-trivial relationships between these parameters, supporting the development of a slice efficiency index that balances resource allocation against transmission reliability.}
        \label{fig:Slice_index}
    \end{minipage}
\end{figure}

\subsection{LLM-Aware Resource Efficiency Index}
Based on Figure.\ref{fig:Compre_heat}, our dataset analysis reveals complex interdependencies between latency, scheduled uplink bytes, and request bytes. The visualization demonstrates non-linear relationships between these critical parameters, suggesting the existence of multi-dimensional optimization boundaries that traditional isolated metrics fail to capture. The color-coded matrix illuminates several key observations:
\begin{enumerate}
    \item Regions of optimal performance emerge at specific combinations of resource allocation and request volume, rather than following a monotonic improvement pattern with increasing resources.
    \item Latency exhibits plateau effects at certain throughput thresholds, indicating saturation points where additional resources yield diminishing returns.
    \item Request volume influences the relationship between resource allocation and latency in non-trivial ways, creating complex performance landscapes that require holistic optimization.
\end{enumerate}

It is important to note that our heatmap representation, constrained by dimensionality limitations of visualization, necessarily focuses on three representative parameters. These selected metrics—latency, scheduled uplink bytes, and request bytes—serve as proxies for more complex system characteristics. For instance, scheduled uplink bytes approximate but do not completely capture overall resource utilization patterns across the network stack, while request bytes provide insight into workload demands without fully representing request resolution or complexity variations. Nevertheless, these parameters offer meaningful approximations of the fundamental system dynamics at play.

To quantify these relationships, we propose the \textbf{LLM-Aware Resource Efficiency Index (LAREI)}, formulated as:
\[
\text{LAREI} = \frac{\text{RDV} \cdot \log(1 + \text{LLM\_Para})}{\text{Resources} \cdot \text{Latency}} \cdot \omega
\]
Where:
\begin{itemize}
    \item \textbf{RDV (Request Data Volume)} quantifies the amount of data requested by the user (approximated by request bytes in our dataset)
    \item \textbf{LLM\_Para (LLM Parameter Count)} represents the computational complexity of the language model (measured in billions of parameters)
    \item \textbf{Resources (Allocated Communication Resources)} represents the network resources provisioned for transmission (approximated by scheduled uplink bytes)
    \item \textbf{Latency} measures end-to-end response time
    \item $\omega$ is a normalization coefficient that adjusts for system-specific characteristics
\end{itemize}

This formulation captures the fundamental efficiency relationship observed in our dataset while incorporating the computational dimension of LLM services. The logarithmic scaling of the parameter count acknowledges the empirically observed sub-linear relationship between model size and perceived quality improvements. This multidimensional index effectively balances the communication-computation trade-off inherent in LLM service delivery over wireless networks.

The LAREI serves as a comprehensive metric for evaluating resource allocation algorithms specifically tailored to LLM communication services, enabling researchers to identify optimal operating configurations that balance communication efficiency with model sophistication.

\subsection{LLM Slice Efficiency Quotient}
Figure.~\ref{fig:Slice_RB}, Figure~\ref{fig:Slice_bytes} and Figure.~\ref{fig:Slice_index} reveal critical insights into network slicing performance that challenge conventional resource allocation assumptions. It demonstrates that higher resource allocation does not necessarily translate to proportionally lower error rates, suggesting an efficiency frontier in slice optimization. The comparative visualization of BLER between slice-enabled and normal traffic illustrates that excessive resource provisioning can potentially increase transmission errors.

Figure.~\ref{fig:Slice_index} further illuminates this phenomenon through a correlation heatmap between slice configuration, request bytes, and uplink BLER. The visualization reveals distinct patterns where certain slice configurations exhibit superior error performance despite lower resource allocation, while others show degraded performance despite abundant resources. This empirical evidence supports the development of a slice efficiency metric that transcends simple resource allocation quantities.

It should be emphasized that our heatmap analysis, while instructive, represents a dimensional reduction of a multifaceted system. The three parameters visualized—slice configuration, request bytes, and BLER—function as indicative proxies for the broader system dynamics. For example, slice configuration encapsulates numerous implementation details not fully represented in the visualization, while BLER provides insight into transmission reliability without comprehensively capturing all aspects of service quality. Despite these limitations, the observed patterns reveal fundamental relationships that inform our metric development.

Based on these observations, we propose the \textbf{LLM Slice Efficiency Quotient (LSEQ)}:
\[
\text{LSEQ} = \frac{\text{RDV\_slice} \cdot (1 - \text{Error Rate}) \cdot \sqrt{\text{LLM\_Para\_slice}}}{\text{Slice Resources}} \cdot \delta
\]
Where:
\begin{itemize}
    \item \textbf{RDV\_slice} represents the amount of data requested by users within the slice
    \item \textbf{Error Rate} quantifies transmission errors (represented by BLER in our dataset)
    \item \textbf{LLM\_Para\_slice} reflects the computational complexity of the language model deployed within the slice
    \item \textbf{Slice Resources} measures the communication resources provisioned to the slice
    \item $\delta$ is a calibration parameter that normalizes across different deployment scenarios
\end{itemize}

This formulation integrates the computational dimension of LLM services into the slice efficiency evaluation, recognizing that slices supporting more complex models (with higher parameter counts) deliver potentially higher service value. The square root scaling of parameter count reflects the diminishing returns observed in perceived quality as model size increases. The multiplicative error factor $(1 - \text{Error Rate})$ ensures that transmission reliability appropriately influences the efficiency evaluation, consistent with our empirical observations.

The LSEQ enables researchers to evaluate slicing algorithms specifically designed for LLM service delivery, accounting for the complex interplay between resource allocation, transmission reliability, and model sophistication. It provides a theoretically sound framework for optimizing slice configurations in LLM-oriented wireless communication systems.
\begin{onecolumn}
\section{All Matrics of Dataset}
\label{appendixMatrics}
\definecolor{layerblue}{RGB}{173,216,230}
\newcolumntype{y}{>{\raggedright\arraybackslash}p{3.5cm}}
\begin{table*}[htbp]
    \centering
    \caption{UE Layer Dataset Metrics}
    \label{tab:ue_metrics}
    \footnotesize
    \setlength{\tabcolsep}{3pt}
    \renewcommand{\arraystretch}{1.25}
    \begin{tabular}{@{}p{1.8cm}yp{7.2cm}@{}}
        \toprule
        \textbf{Layer} & \textbf{Metric} & \textbf{Description} \\ 
        \midrule
        
        \cellcolor{layerblue}UE
        & Timestamp & - Request initiation timestamp (Unix epoch milliseconds) \\
        & Wireless Comm Time &- UE-gNB air interface duration (PDCP layer measurement) \\
        & Total Comm Time &- UE-side end-to-end latency (request to response) \\
        & Tx Image Resolution &- Original image dimensions (e.g., 1920×1080) \\
        & Rx Image Resolution &- Received image dimensions post-processing \\
        & Expected Word Count &- User-specified desired explanation length \\ 
        & Actual Word Count &- LLM-generated response word count \\
        & LLM Model &- Model architecture (LLaMA, LLaVA, etc.) \\
        & Request Mode &- \texttt{image\_request} or \texttt{text\_request} \\
        & Upload Periodicity &- Uplink interval (ms, 0=event-driven) \\
        & Uplink Time &- UE-to-network latency (RLC layer) \\
        & Downlink Time &- Network-to-UE latency (PDCP layer) \\
        & Downlink Text Size &- Response payload size (UTF-8 bytes) \\
        & Uplink Bytes &- Total bytes of the request (uplink) \\
        & Downlink Bytes &- Total bytes of the response (downlink) \\
        \bottomrule
    \end{tabular}
\end{table*}

\begin{table*}[htbp]
    \centering
    \caption{Edge Server Layer Dataset Metrics}
    \label{tab:server_metrics}
    \footnotesize
    \setlength{\tabcolsep}{3pt}
    \renewcommand{\arraystretch}{1.25}
    \begin{tabular}{@{}p{1.8cm}yp{7.2cm}@{}}
        \toprule
        \textbf{Layer} & \textbf{Metric} & \textbf{Description} \\ 
        \midrule
        
        \cellcolor{layerblue}Server
        & LLM Inference Time &- Model forward pass latency (ms) \\
        & Server Processing Time &- Processing time in server \\
        & Input Tokens &- Tokens consumed by LLM \\
        & Output Tokens &- Tokens generated by LLM \\
        & Cold Start Time &- Disk-to-RAM loading time \\
        & Warm Start Time &- Cache reload time \\
        & BLEU Score &- Text quality evaluation metric \\
        & ROUGE Score &- Recall-based text evaluation \\
        & Semantic Score &- Embedding similarity score \\
        & GPU Utilization &- GPU compute usage (\%) \\
        & VRAM Usage &- Video memory consumption \\
        & Downlink Image &- Base64 output image \\
        & Response Text &- LLM-generated response \\
        \bottomrule
    \end{tabular}
\end{table*}

\begin{table*}[htbp]
    \centering
    \caption{RAN Layer Dataset Metrics}
    \label{tab:ran_metrics}
    \footnotesize
    \setlength{\tabcolsep}{3pt}
    \renewcommand{\arraystretch}{1.25}
    \begin{tabular}{@{}p{1.8cm}yp{7.2cm}@{}}
        \toprule
        \textbf{Layer} & \textbf{Metric} & \textbf{Description} \\ 
        \midrule
        
        \cellcolor{layerblue}RAN
        & gNB Timestamp &- gNB timestamp alignment (Unix epoch ms) \\
        & Frame Number &- 5G NR system frame (0-1023) \\
        & Slot Number &- Within-frame slot index (0-159) \\
        & IMSI &- International Mobile Subscriber Identity \\
        & RNTI &- Radio Network Temporary Identifier \\
        & UE ID &- Device identifier within gNB \\
        & UE Number &- Logical connection index \\
        & DL Throughput &- Instantaneous downlink rate (Mbps) \\
        & UL Throughput &- Instantaneous uplink rate (Mbps) \\
        & PH (dB) &- UE Power Headroom report \\
        & PCMAX (dBm) &- Max transmission power \\
        & Avg RSRP &- Reference Signal Received Power \\
        & CQI &- Channel Quality Indicator (0-15) \\
        & RI &- MIMO Rank Indicator \\
        & DL MCS &- Downlink Modulation and Coding Scheme \\
        & UL MCS &- Uplink Modulation and Coding Scheme \\
        & Scheduled UL Bytes &- Uplink buffer status \\
        & Estimated UL Buffer &- gNB buffer estimation (bytes) \\
        & DL PDUs Total &- Aggregated downlink PDUs \\
        & DL BLER &- Downlink Block Error Rate (\%) \\
        & UL BLER &- Uplink Block Error Rate (\%) \\
        & DL-SCH Bytes &- Downlink shared channel payload \\
        & DL-SCH RBs &- Downlink Resource Blocks allocated \\
        & UL-SCH Bytes &- Uplink shared channel payload \\
        & UL-SCH RBs &- Uplink Resource Blocks allocated \\
        & UL MAC SDUs &- MAC Service Data Units count \\
        & Primary Slice Max &- Main slice maximum resources \\
        & Primary Slice Min &- Main slice minimum guarantee \\
        & Secondary Slice Max &- Secondary slice maximum \\
        & Secondary Slice Min &- Secondary slice minimum \\
        \bottomrule
    \end{tabular}
\end{table*}
\end{onecolumn}

\end{document}